\documentclass[journal]{IEEEtran}
\usepackage{cite}

\ifCLASSINFOpdf
  \usepackage[pdftex]{graphicx}
\else
\fi
\usepackage{amsmath}
\usepackage{amsfonts}
\usepackage{amssymb}

\usepackage{array}

\usepackage[dvipsnames]{xcolor}
\usepackage{hyperref}
\usepackage{amsthm}
\theoremstyle{plain}

\usepackage{textcomp}
\usepackage{bbm}
\usepackage{algorithm}
\usepackage{algpseudocode}
\usepackage{tikz}
\usetikzlibrary{arrows.meta}
\usepackage[font=footnotesize]{caption}

\definecolor{localization}{RGB}{50,168,82}
\definecolor{data}{RGB}{77,123,232}

\makeatletter
\def\@makefnmark{%
  \leavevmode
  \raise.9ex\hbox{\fontsize\sf@size\z@\normalfont\tiny\@thefnmark}}
\makeatother

\usepackage{glossaries}

\newacronym{3gpp}{3GPP}{3rd Generation Partnership Project}
\newacronym{bs}{BS}{base station}
\newacronym{cdf}{CDF}{cumulative distribution function}
\newacronym{pdf}{PDF}{probability density function}
\newacronym{csi}{CSI}{channel state information}
\newacronym{ue}{UE}{user equipment}
\newacronym{los}{LoS}{line-of-sight}
\newacronym{nlos}{NLoS}{non-line-of-sight}
\newacronym{mdt}{MDT}{minimization of drive tests}
\newacronym{ofdm}{OFDM}{orthogonal frequency division multiplexing}
\newacronym{tdoa}{TDOA}{time difference of arrival}
\newacronym{toa}{TOA}{time of arrival}
\newacronym{urllc}{URLLC}{ultra-reliable low-latency communications}
\newacronym{mar}{MAR}{maximum achievable rate}
\newacronym{pcr}{PCR}{probably correct reliability}
\newacronym{snr}{SNR}{signal-to-noise ratio}
\newacronym{peb}{PEB}{position error bound}
\newacronym{fi}{FI}{Fisher information}
\newacronym{fim}{FIM}{Fisher information matrix}
\newacronym{crlb}{CRLB}{Cram\'{e}r-Rao lower bound}
\newacronym{6g}{6G}{sixth-generation wireless}
\newacronym{5g}{5G}{fifth-generation wireless}
\newacronym{iot}{IoT}{internet of things}
\newacronym{ar}{AR}{augmented reality}
\newacronym{mtc}{MTC}{machine-type communication}
\newacronym{ckm}{CKM}{channel knowledge map}
\newacronym{ran}{RAN}{radio access network}
\newacronym{mimo}{MIMO}{multiple input multiple output}

\newcommand{\mbf}[1]{\mathbf{#1}}
\newcommand{\mbs}[1]{\boldsymbol{#1}}
\newcommand{\mR}{\mathbb{R}}
\newcommand*{\cond}{\hspace*{1pt} |\hspace*{1pt}}
\newcommand{\norm}[1]{\left\lVert#1\right\rVert}
\newcommand{\mC}{\mathbb{C}}
\newcommand{\T}{\textsf{T}}
\renewcommand{\H}{\textsf{H}}

\DeclareMathOperator{\Var}{Var}

\newcommand{\picscale}{0.55}

\begin{document}
\title{On the Statistical Relation of Ultra-Reliable Wireless and Location Estimation}

\author{Tobias~Kallehauge,
        Martin~Voigt~Vejling,
        Pablo~Ramírez-Espinosa,
        Kimmo~Kansanen,~Henk~Wymeersch~\IEEEmembership{Senior Member,~IEEE}, and Petar~Popovski,~\IEEEmembership{Fellow,~IEEE}% <-this % stops a space
\thanks{This work was supported in part by the Villum Investigator Grant “WATER” from the Velux Foundation, Denmark. The work by P. Ram\'irez-Espinosa has been supported by a ``Mar\'ia Zambrano" Fellowship funded by the European Union – Next Generation EU via the Ministry of Universities of the Spanish Government.}
\thanks{T. Kallehauge, M. V. Vejling, and P. Popovski are with the Department of Electronic Systems, Aalborg University, Denmark (e-mail: tkal@es.aau.dk; mvv@es.aau.dk; petarp@es.aau.dk).}% <-this % stops a space
\thanks{P. Ramírez-Espinosa is with Department of Signal Theory, Networking and Communications, Research Centre for Information and Communication Technologies (CITIC), University of Granada, Spain (e-mail: pre@ugr.es)}%
\thanks{K. Kansanen is with the Norwegian University of Science and Technology, Trondheim, Norway (e-mail: kimmo.kansanen@ntnu.no).}%
\thanks{H. Wymeersch is with the Department of Electrical Engineering, Chalmers University of Technology, Gothenburg, Sweden (e-mail: henkw@chalmers.se).}}



\maketitle

\begin{abstract}
Location information is often used as a proxy to guarantee the performance of a wireless communication link. However, localization errors can result in a significant mismatch with the guarantees, particularly detrimental to users operating the ultra-reliable low-latency communication (URLLC) regime. This paper unveils the fundamental statistical relations between location estimation uncertainty and wireless link reliability, specifically in the context of rate selection for ultra-reliable communication. We start with a simple one-dimensional narrowband Rayleigh fading scenario and build towards a two-dimensional scenario in a rich scattering environment. The wireless link reliability is characterized by the meta-probability, the probability with respect to localization error of exceeding the outage capacity, and by removing other sources of errors in the system, we show that reliability is sensitive to localization errors. The $\epsilon$-outage coherence radius is defined and shown to provide valuable insight into the problem of location-based rate selection. However, it is generally challenging to guarantee reliability without accurate knowledge of the propagation environment. Finally, several rate-selection schemes are proposed, showcasing the problem's dynamics and revealing that properly accounting for the localization error is critical to ensure good performance in terms of reliability and achievable throughput.
\end{abstract}

% \begin{IEEEkeywords}

% \end{IEEEkeywords}

% \fi
%\IEEEpeerreviewmaketitle

\section{Introduction} \label{sec:intro}
\IEEEPARstart{A}{}s wireless communication moves towards developing and deploying the \gls{6g} networks, localization and sensing are expected to play a crucial role in shaping the future of communication technology. This includes the area of \gls{urllc}, where \gls{6g} aims to provide unprecedented connectivity, low latency, and reliable communication for a wide range of applications, including autonomous vehicles, industrial automation, \gls{ar}, and more. Key enablers for high-accuracy localization include using new and higher frequency bands, deploying intelligent surfaces in the propagation environment, intelligent beamforming, and applying machine learning and artificial intelligence techniques \cite{Lima2021Sensing}. The amount of available data is also increasing with, %We also see a boom in the availability of data, 
e.g., the recent standardization by the \gls{3gpp} of \gls{mdt}\cite{3gpp_MDT}, allowing the operators to utilize the end-user devices for collecting location-specific measurements. Furthermore, the concept of \textit{channel charting} has been shown to provide location-like information by exploiting rich \gls{csi} samples from massive \gls{mimo} systems, thus providing an attractive replacement for localization \cite{Ferrand2021Chart}. With  the increasing availability of user measurements associated with high-accuracy location information, it becomes progressively more viable to use the location as a proxy for local channel conditions. Examples of this include using location information to predict traffic levels in \gls{mtc} \cite{Shoaei2021traffic}, location-aided beamforming for vehicle communication \cite{Garcia2016}, and a protocol for scheduling time-frequency resources in RIS-aided OFDM systems based on localization information \cite{saggese2023localization}.
%and prediction of path loss levels based on coverage maps \cite{predict_link}. 
Related to these ideas is the concept of \textit{\glspl{ckm}} --- a database containing various channel-related information tagged with associated location information. \Glspl{ckm} are thus expected to play a vital role in enabling environment-aware communication in \gls{6g}\cite{Zeng2021knowledge}.

Concerning \gls{urllc} in specific, localization can be exploited by using that reliability of wireless transmission is related to, among others, the behavior of the propagation channel, which is inherently correlated with spatial location. Consequently, exploiting this relationship is envisioned as a promising direction for \gls{urllc} \cite{risk_URLLC, Yoon2021rate, kallehaugeGlobecom2022}. In \cite{risk_URLLC}, it is proposed to use a user-generated radio map at the \gls{ran} control center to estimate the risk of outages and determine network configurations. In \cite{Yoon2021rate}, a rate adjustment scheduler using location-aware prediction of interference is proposed for \gls{urllc} networks. Similarly, \cite{kallehaugeGlobecom2022} uses location information for a rate selection scheme based on a radio map to enable statistical guarantees for \gls{urllc}. These examples differ from conventional approaches, where samples are acquired over time to estimate channel statistics and, thus, reliability \cite{marko}. Considering the latency introduced by estimating channel statistics, a communication system that predicts reliability based on localization using only a few measurements is an attractive alternative. 
An important aspect of these location-based inference strategies is the capability to estimate the location accurately. The literature rarely addresses this issue, and it is common to assume error-free localization. This practice is not without justification; indeed, \gls{6g} networks are expected to have centimeter-level localization accuracy\cite{Zeng2021knowledge}, but these metrics are yet to be realized with current performances in the meter-level range \cite{xiao2022overview}. Regardless, the accuracy of localization could be a liability for systems in the ultra-reliable domain relying on location information, which raises the fundamental question: \emph{How can the accuracy of the localization procedures impact the wireless reliability guarantees?}

We posed this question and provided the first preliminary study in \cite{kallehaugeWCL2022}, where the basic relations between location uncertainty and reliability were analyzed based on a simple channel model in a one-dimensional scenario. In this work, we generalize the analysis to a broader range of scenarios, including two-dimensional localization and more realistic channels with a variable number of paths. The general setup is the following: A \gls{ue} in the \gls{urllc} regime communicates with a \gls{bs} and should choose the transmission parameters to fulfill certain requirements in terms of reliability. It is assumed that estimation of \gls{csi} is not possible, e.g., in a scenario with aperiodic and spontaneous transmissions where the latency constraint prohibits \gls{csi} acquisition. The \gls{ue} must therefore rely on statistical information of long-term channel conditions to ensure reliability \cite{wireless_access}. We here analyze the feasibility of a localization-based approach, where the location is used as a proxy for characterizing the channel. To isolate the impact of location uncertainty on reliability, this paper assumes that the propagation environment has already been mapped, e.g., using a \gls{ckm}, such that the channel statistics are perfectly known at each location. Thus, if the location were also perfectly known, the \gls{ue} would correctly allocate resources to guarantee a target level of reliability, but given the uncertainty of the estimated location, the predicted reliability is also uncertain. The \gls{ue} must therefore account for scenarios where, e.g., the signal level is weaker at the true location than at the estimated location. With this setup, this paper will study the problem of location-based rate selection and how localization error can affect the resulting reliability and throughput. Reliability is statistically characterized following the \textit{probably correct reliability} approach introduced in \cite{marko}, and localization performance is modeled, among others, through \gls{fi} analysis \cite{equivalent_fisher}. The contributions of this paper are  summarized in the following: 
\begin{itemize}
    \item Two scenarios and propagation environments are analyzed, starting with a simple one-dimensional scenario subjected to narrowband Rayleigh fading, and continuing with a wideband channel with propagation conditions according to standardized \gls{3gpp} models generated using the simulation tool \textit{QuaDRiGa}\cite{Quadriga}. Significant contributions compared to \cite{kallehaugeWCL2022} include closed-form solutions for rate selection in the Rayleigh fading case, analysis of two-dimensional scenarios, and more realistic channel models, including large-scale fading effects such as shadowing.  
    \item The relation between local variations of channel statistics and the performance of location-based rate selection is studied extensively. Specifically, we explore how the \textit{$\epsilon$-outage coherence radius} is a good performance indicator. The impact of the localization error severity is also thoroughly evaluated. 
    \item Several location-based rate selection schemes that account for localization uncertainty are considered. The reliability and throughput of these schemes are then studied both analytically and through simulation. For simple environments (narrowband Rayleigh fading), we analytically solve the rate selection problem, and numerical methods are used in the wideband scenario.
    \item By analyzing the performance of the proposed schemes, it is highlighted that the efficiency of location-based rate selection schemes --- i.e., spectral efficiency for a given reliability target --- is severely compromised if location errors are not properly considered.
\end{itemize}
The remainder of the paper is structured as follows. Section \ref{sec:reliability_and_localization} introduces the general channel model and formally introduces the problem of location-based rate selection. Section \ref{sec:narrowband_rayleigh} studies a simple Rayleigh fading channel model with average received power according to path loss and one-dimensional localization. 
Section \ref{sec:wideband} examines the problem in a two-dimensional wideband scenario based on numerical methods, and the paper is concluded in section \ref{sec:conclusion}. Error bounds of \gls{toa}-based localization are derived in App. \ref{sec:localization}. 

\emph{Supplementary resources:} The code used for simulations and figures shown in the paper can be found at \href{https://github.com/TobiasKallehauge/Localization-and-Reliability-in-URLLC}{https://github.com/TobiasKallehauge/Localization-and-Reliability-in-URLLC}.

\emph{Notation:} $\mathbb{N}$, $\mathbb{R}$, $\mR_{+}$, $\mathbb{C}$ denotes the sets of positive integers, real numbers, non-negative real numbers, and complex numbers, respectively. $\mathfrak{R}(z)$ and $\mathfrak{I}(z)$ are the real and imaginary parts of complex number $z$, and $\jmath$ is the imaginary unit. $(\cdot)^\T$ and $(\cdot)^\H$ are the matrix transpose and conjugate transpose, respectively, and $\norm{\cdot}$ is the $\ell_2$-norm. $\mathcal{N}(\mu,\sigma^2)$ and $\mathcal{CN}(\mu,\sigma^2)$ denote Gaussian and circularly symmetric complex Gaussian distributions with mean $\mu$ and variance $\sigma^2$. $\ln$ is used for the natural logarithm, while logarithms with other bases will be denoted explicitly in the subscript, e.g., $\log_2$.  The $n\times n$ identity matrix is denoted $\mbf{I}_n$. Finally, $E[\cdot]$, $\Var[\cdot]$, $\mathbb{T}[\cdot]$, denote, respectively, the expectation, variance, and trace operators.

\section{Location-Based Rate Selection} \label{sec:reliability_and_localization}
This section formally introduces the problem of location-based rate selection. We start by introducing the system and channel model used throughout the paper. 

\subsection{Communication System and Channel Model} \label{sec:model}
A cellular network with an arbitrary number of \glspl{bs} supports both localization and ultra-reliable communications. The \glspl{bs} and the \glspl{ue} are all equipped with a single antenna, and an \gls{ofdm} modulation scheme is considered transmitting at bandwidth $W$ over $N$ subcarriers spaced $\Delta_f = W/N$ for $N \geq 1$. Thus, with normalized symbol $\mbf{s} \in \mC^N$ (i.e., $E[\norm{\mbf{s}}^2] = N$), the received baseband signal from a \gls{bs} in the frequency domain at subcarrier $j$ is given by\footnote{We assume identical uplink and downlink channels. Note the assumption of no intersymbol interference due to \gls{ofdm} modulation.} %The communication channel is modeled for a \gls{ue} at location $\mbf{x} \in \mathcal{R}$ and one of the \glspl{bs} at location $\mbf{x}_{\text{bs}} \in \mathcal{R}$, where $\mathcal{R} \subset \mR^D$ is the cell. We analyze the two cases $D = 1$ and $D = 2$ in the paper. We model the channel as a tapped delay line channel where $K$ discrete paths arrive at different delays. 
\begin{align} \label{eq:sysmod}
y_{j} =\sqrt{P_{\text{tx}}}\sum_{k=1}^K a_k d_j(\tau_k) s_j + n_{j},\quad  j = 0, \dots, N - 1
\end{align}
where $P_{\text{tx}}$ is the transmit power per subcarrier, $K$ is the number of paths or clusters, $n_{j}\sim \mathcal{CN}(0,  W N_0)$ is the noise term with variance $WN_0$, and $N_0$ is the power spectral density of the noise. The channel within each path/cluster is defined by the delay $\tau_k\in\mathbb{R}_+$ and the coefficient $a_k \in \mathbb{C}$, where $d_j(\tau_k) = \exp(-2\pi\jmath \Delta_f j \tau_k)$ is the induced phase rotation due to the signal delay. With the possible exception of a \gls{los} path, $a_k$ for $k = 1,\dots, K$, are modeled as stochastic variables arising from the sum of multiple --- irresolvable --- scattered paths within each cluster. Note that the channel coefficients $a_k$ and delays $\tau_k$ for clusters $k = 1,\dots, K$ depend on both the \gls{ue} location $\mbf{x} \in \mathcal{R}$ and the \gls{bs} location $\mbf{x}_{\text{bs}} \in \mathcal{R}$ with $\mathcal{R} \subset \mR^D$, but this dependence is omitted for the clarity of presentation. 

To study the reliability of location-based rate selection, we consider the following two-step protocol:
\begin{enumerate}
 \item A \gls{ue} at location $\mbf{x}$ estimates its location as $\hat{\mbf{x}}$ using either \gls{toa}-based localization by connecting to at least $D+1$ \glspl{bs} or by any other external system. The model for the localization error will depend on the scenario, as specified later. 
\item Using the estimated location $\hat{\mbf{x}}$, the \gls{ue} selects its transmission rate, denoted as $R(\hat{\mbf{x}})$. It then starts the communication with just one of the \glspl{bs} by sending data over the channel in \eqref{eq:sysmod}.
\end{enumerate}
As an important remark, note that the channels used in steps 1 and 2 do not need to be the same, as exemplified in Sec. \ref{sec:narrowband_rayleigh}, where localization uses a separate system.

\subsection{Communication Reliability}

At the physical layer, we focus on the outage probability as a reliability metric, which characterizes the event when the selected rate exceeds the channel capacity. With maximum ratio combining at the receiver, the normalized capacity of the channel between a \gls{bs} and the \gls{ue} at location $\mbf{x}$ becomes
\begin{align}
    C(\mbf{x}) = \sum_{j=0}^{N - 1} \log_2\left(1 + \frac{P_{\text{tx}}}{W N_0} |h_j|^2\right) , \label{eq:capacity}
\end{align}
where $h_j = \sum_{k=1}^K a_k d_j(\tau_k)$ such that the random variation of $a_k$ induces a random variation in the channel capacity. The outage probability for estimated location $\hat{\mbf{x}}$ and selected rate $R(\hat{\mbf{x}})$ is then denoted \cite{Durisi2016short}
\begin{align}
    p_{\text{out},\mbf{x}}(R(\hat{\mbf{x}})) = P(C(\mbf{x}) \leq R(\hat{\mbf{x}})).
\end{align}
Given a target level of reliability $\epsilon \in (0,1)$, we require that $p_{\text{out},\mbf{x}}(R(\hat{\mbf{x}})) \leq \epsilon$ for data transmission. Hence, if the statistics of $C(\mbf{x})$ were known, this requirement would be fulfilled by selecting
\begin{align}
    R(\mbf{x}) = C_{\epsilon}(\mbf{x}) = \sup \{R\geq 0 \cond p_{\text{out},\mbf{x}}(R) \leq \epsilon\},
\end{align}
known as the \emph{$\epsilon$-outage capacity}. However, in practice, the system must rely on the available information to estimate the $\epsilon$-outage capacity \cite{marko, kallehauge2023magazine}. As mentioned, we will study the case of using location as a proxy to select the rate, assuming that $C_{\epsilon}(\mbf{x})$ is known for every $\mbf{x}$ in the cell $\mathcal{R}$ such that the only source of error is the localization procedure. Naturally, this represents an ideal scenario but allows us to extract insights and analyze the feasibility of location-based rate selection approaches (also providing an upper bound in performance). To account for the location uncertainty, we focus on the concept of \textit{meta-probability}, defined as the probability that the target reliability $\epsilon$ is exceeded, i.e., \cite{marko} 
\begin{align}
    \tilde{p}_{\epsilon}(\mbf{x}) = P_{\hat{\mbf{x}}}(p_{\text{out},\mbf{x}}(R(\hat{\mbf{x}})) > \epsilon) = P_{\hat{\mbf{x}}}(R(\hat{\mbf{x}}) > C_{\epsilon}(\mbf{x})),
\end{align}
where the outer probability is with respect to the estimated location $\hat{\mbf{x}}$. Reliability is thus targeted by designing the rate selection function $R$ such that the meta-probability does not exceed some target $\delta \in (0,1)$, referred to as the \textit{confidence parameter}. Since the meta-probability can be arbitrarily low if a very conservative rate is selected, we combine this performance indicator with the \textit{throughput ratio}, defined as \cite{marko, kallehaugeGlobecom2022}
\begin{align}
    \overline{T}(\mbf{x}) = \frac{E_{\hat{x}}[R(\hat{\mbf{x}})(1 -p_{\text{out},\mbf{x}}(R(\hat{\mbf{x}}))]}{C_{\epsilon}(\mbf{x})(1-\epsilon)}, \label{eq:Througput_ratio}
\end{align}
which quantifies the penalty introduced by the location uncertainty. The throughput ratio is exactly one if the location is perfectly known and, hence, $R(\hat{\mbf{x}}) = C_{\epsilon}(\mbf{x})$.

The problem of location-based rate selection thus reduces to learning a rate selection function $R : \mathcal{R}\to \mathbb{R}_+$, which maximizes the --- average --- throughput ratio while ensuring that the meta-probability stays below the target, i.e.,
\begin{align}
    \max_{R : \mathcal{R}\to \mathbb{R}_+}  \frac{1}{|\mathcal{R}|} \int_{\mbf{x} \in \mathcal{R} }\overline{T}(\mbf{x}) \ d\mbf{x}\quad \text{st.} \quad \tilde{p}_{\epsilon}(\mbf{x}) \leq \delta \ \forall \mbf{x} \in \mathcal{R}. \label{eq:opmization}
\end{align}
The above problem is non-trivial, and different heuristic solutions will be presented and compared throughout this paper.

\section{Rate Selection with Narrowband Rayleigh Fading Channel} \label{sec:narrowband_rayleigh}
We start with laying the foundation of the location-based rate selection problem by studying a simple, albeit illustrative, scenario. This allows tractable analysis of the problem, thus greatly improving the understanding acquired in \cite{kallehaugeWCL2022}, where the insight was mainly based on numerical simulations. We consider the case of narrowband Rayleigh channel --- a special case of \eqref{eq:sysmod} --- with the added simplification of a one-dimensional cell ($D = 1$). While localization is often based on wideband systems, this is not necessarily incompatible with the scenario here. Practical examples could include narrowband communication with GPS for localization or, as proposed in \cite{Poruia2019Bluetooth}, a phase-based localization approach where channel-hopping is used to sample a wide spectrum to resolve multipath fading (despite each channel being narrowband). 

\subsection{Scenario}
We consider a \gls{bs} located at $x_{\text{bs}} = 0$ communicating with a \gls{ue} at location $x$ within the cell $[x_{\min},x_{\max}]$ with $x_{\min} > 0$. Under the Rayleigh fading assumption, the channel in \eqref{eq:sysmod} reduces to a single \gls{nlos} component as
\begin{align}
    y = \sqrt{P_{\text{tx}}}hs + n, \label{eq:narrowband_channel}
\end{align}
where $h = \sqrt{\overline{P}(x)}\alpha$, $\alpha \sim \mathcal{CN}(0,1)$ is the power-normalized fading coefficient and $\overline{P}(x)$ is the average received power modeled according to the path loss formula \cite[ch. 2]{goldsmith2005wireless}
\begin{align}
     \overline{P}(x) = G_0x^{-\eta}, \ x > 0 \label{eq:PL}
\end{align}
with $G_0 \in \mR$ a gain factor and $\eta > 0$ the path loss exponent. The average \gls{snr} is denoted $\overline{\gamma}(x) = \gamma_0\overline{P}(x)$, where $\gamma_0 = \frac{P_{\text{tx}}}{WN_0}$, such that the instantaneous \gls{snr} is $\gamma(x) = \overline{\gamma}(x)|\alpha|^2$, which follows an exponential distribution with mean $\overline{\gamma}(x)$. 

We further assume an unbiased location estimator with normally distributed errors (which is consistent with the localization theory presented in App. \ref{sec:localization} and used in the next section). Hence, the estimated location is modeled as $\hat{x} \sim \mathcal{N}(x,\sigma_{x}^2)$, where $x$ is the true (unknown) \gls{ue} location and $\sigma_{x}^2 \triangleq \Var[\hat{x}]$ models the localization variance for each $x \in [x_{\min},x_{\max}]$. While not necessary for analyzing the problem, it is assumed that $P(\hat{x} < 0) = 0$, which simplifies the analysis by removing some edge cases where the estimated location is to the left of the \gls{bs}.

In this case, the capacity in \eqref{eq:capacity} reduces to $C(x) = {\log_2(1 + \overline{\gamma}(x)|\alpha|^2)}$, yielding 
\begin{align}
    P(C(X) \leq R) &=  P\left(|\alpha|^2 \leq \frac{2^R-1}{\overline{\gamma}(x)}\right) \\
     &= 1 - \exp\left(-\frac{2^R-1}{\overline{\gamma}(x)} \right) \label{eq:CDF_rayleigh}
\end{align}
for $R > 0$, where the last equality follows from $|\alpha|^2$ being expontentially distributed with unit mean. The $\epsilon$-outage capacity then follows directly from \eqref{eq:CDF_rayleigh} as
\begin{align}
    C_{\epsilon}(x) = \log_2(1 - \overline{\gamma}(x)\ln(1-\epsilon)). \label{eq:rayleigh_outage}
\end{align}
Note that $C_{\epsilon}$ is a smooth function of $x$ due to path loss being the only large-scale fading phenomenon in the model, so the scenario here does not capture effects like blockages that cause an abrupt change in the channel. Lastly, we introduce the \emph{$\epsilon$-outage coherence radius} (or simply coherence radius) defined as \textit{the minimum radius from the \gls{ue} location $x$ where the maximum relative change in $C_{\epsilon}$ within the radius exceeds some threshold $t > 0$}, i.e.,
\begin{align}
    \text{CR}(x) = \min_{\varrho > 0}\left\{ \varrho \hspace*{1pt} \left|\hspace*{1pt} \max_{z \in [x-\varrho,x+\varrho]} \left\{\frac{|C_{\epsilon}(x) - C_{\epsilon}(z)|}{C_{\epsilon}(x)}\right\} > t \right. \right\}. \label{eq:coherence_1d}
\end{align}
The coherence radius is thus a measure of the local spatial variation of the channel conditions and will be useful to get insight into the location-based rate selection problem. It can be shown using \eqref{eq:rayleigh_outage} that\footnote{The approximation in \eqref{eq:coherence_analytical} is based on a 1st order Taylor expansion that is accurate whenever $\gamma_0G_0\ln(1-\epsilon) \approx 0$. The approximation errors with the settings used in this section are less than $10^{-4}$ m for all $x \in [20,100]$ m .}
\begin{align}
    \text{CR}(x) &= x - \left(\frac{\gamma_0G_0\ln(1-\epsilon)}{1 - \left(1 - \gamma_0G_0x^{-\eta}\ln(1-\epsilon)\right)^{1+t}} \right)^{1/\eta} \\
    &\approx x(1-(1+t)^{-1/\eta}), \label{eq:coherence_analytical}
\end{align}
being thus approximately proportional to $x$ (i.e., the distance to the \gls{bs}).
\subsection{Backoff Rate Selection}
The structure of \eqref{eq:rayleigh_outage} inspires the trivial solution to \eqref{eq:opmization} 
\begin{align}
R(\hat{x}) = \log_2(1 - \beta\overline{\gamma}(\hat{x})\ln(1-\epsilon)),   \label{eq:rate_select_general_1d_no_shadow}
\end{align}
for some $\beta \in (0,1]$. We denote this solution as \textit{backoff rate selection}, where the rate is chosen as the $\epsilon$-outage capacity at the estimated location $\hat{x}$, but re-scaling the average \gls{snr} by $\beta$ --- backoff --- to account for the uncertainty in the estimated location.
Note the assumption that the average \gls{snr} is known given the location, so the uncertainty in the selected rate is only affected by the localization error, yielding the meta-probability 
\begin{align}
    \tilde{p}_{\epsilon}(x) = P_{\hat{x}}(R(\hat{x}) > C_{\epsilon}(\hat{x})) = P_{\hat{x}}(\beta\overline{\gamma}(\hat{x}) > \overline{\gamma}(x)). \label{eq:meta_1d_noshadow}
\end{align}
It follows that $\beta\overline{\gamma}(\hat{x}) > \overline{\gamma}(x) \Leftrightarrow \hat{x} < x\beta^{1/\eta}$. Defining $\xi = x - \hat{x}$, we see that the outage probability is exceeded whenever $\xi > x(1 - \beta^{1/\eta})$. Since $\xi \sim \mathcal{N}(0,\sigma^2_{x})$, the meta-probability is finally given by
\begin{align}
    \tilde{p}_{\epsilon}(x) &= P_{\xi}(\xi  \geq x(1-\beta^{1/\eta})) \\
    &= 1 - \Phi\left(\frac{x}{\sigma_{x}}(1 - \beta^{1/\eta})\right), \label{eq:meta_backoff}
\end{align}
where $\Phi$ is the \gls{cdf} of the standard normal distribution.

\begin{figure}
    \centering
    \includegraphics[scale = \picscale]{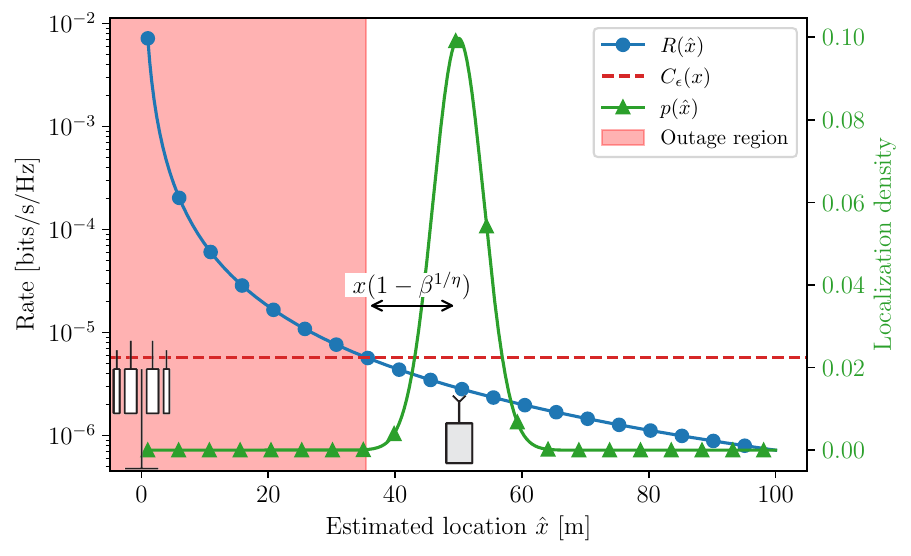}
    \caption{Illustration of backoff rate selection with $\beta = 0.5$ for a user at true location $x = 50$ m with $\sigma_{x}^2 = 16 \text{ m}^2$, $\gamma_0 = 30$ dB, $G_0 = 0$ dB, $\eta = 2$ and $\epsilon = 10^{-5}$. The distance from $x$ to the outage region is $14.6$ m resulting in $\tilde{p}_\epsilon(x) = 0.013~\%$ according to \eqref{eq:meta_backoff}.}
    \label{fig:one_location_1d_no_shadow}
\end{figure}

The performance of the proposed backoff rate selection is exemplified in Fig.~\ref{fig:one_location_1d_no_shadow}, which shows the selected rate $R(\hat{x})$ at different locations and $C_{\epsilon}(x)$ for the true location $x$, together with the localization \gls{pdf}. As predicted by \eqref{eq:rate_select_general_1d_no_shadow}, we see that the selected rate is a decreasing function on $\hat{x}$ since, the larger the distance to the \gls{bs}, the larger the path loss. The figure also depicts the \textit{outage region}, defined as the range of estimated locations where the selected rate exceeds the $\epsilon$-outage capacity, i.e., the region defined by $\xi > x(1-\beta^{1/\eta})$ as discussed previously. Fig.~\ref{fig:one_location_1d_no_shadow} reveals the fundamental observation that the \gls{ue} will choose an overly optimistic rate if it thinks it has better channel conditions than it actually has --- that is, closer to the \gls{bs} than it actually is. Conversely, if the \gls{ue} underestimates the channel conditions ($\hat{x} > x$), it chooses an overly conservative rate, lowering the spectral efficiency. Overall, lower values of $\beta$ lead to more conservative rates and move the outage region further away from the \gls{ue}. These observations are similar to those extracted in \cite{kallehaugeWCL2022}, where reliable transmission can be achieved at the cost of reduced spectral efficiency.

As observed, the value of $\beta$ is critical in the system's performance. Therefore, the problem in \eqref{eq:opmization} is now solved for backoff rate-selection by finding the maximum value of $\beta$ such that  $\tilde{p}_{\epsilon}(x) \leq \delta$ for all $x \in [x_{\min},x_{\max}]$\footnote{The throughput ratio in \eqref{eq:Througput_ratio} increases monotonically for higher selected rates up till the point where the term $1 - p_{\text{out},x}(R(\hat{x}))$ starts to penalize the average rate. When operating the high-reliability domain with $\epsilon \ll 1$, this term is negligible, and the problem in \eqref{eq:opmization} reduces to finding the maximum rates that satisfy the reliability constraint for each location.}. Noting that $\Phi$ in \eqref{eq:meta_backoff} is monotonically increasing, we have that the meta-probability requirement is met by selecting
\begin{align}
    \beta = \left(1 - \frac{\Phi^{-1}(1 - \delta)\sigma_{x^*}}{x^*} \right)^{\eta}, \quad x^* = \min_{x \in [x_{\min},x_{\max}]}\left(\frac{x}{\sigma_{x}}\right), \label{eq:k_analytical}
\end{align}
where $x^*$ is the location experiencing the highest meta-probability, i.e., lowest reliability. Interestingly, it can be proved that $x^*=x_{\min}$ if $\frac{\sigma_{x_{\min}}}{x_{\min}} > \frac{1}{x_0}\displaystyle\int_{0}^{x_0} \Big(\frac{d \sigma_{x_{\min}+x}}{dx}\Big) dx$, for all $x_0\in[0,x_{\max}-x_{\min}]$, which is satisfied when the localization variances grow slower than the distance to the \gls{bs}. The intuition behind why the lowest reliability is experienced close to the \gls{bs} can be understood through the coherence radius. From \eqref{eq:coherence_analytical}, it is apparent that the relative change in channel conditions gradually flattens as the \gls{ue} moves away from the \gls{bs}, and vice versa as the \gls{ue} moves towards the \gls{bs} (as also seen in Fig. \ref{fig:one_location_1d_no_shadow}). So even a small localization error close to a \gls{bs} can cause the \gls{ue} to pick a much higher rate, whereas the same error farther away causes a smaller change in the selected rate. 

\begin{figure}[t]
    \centering
    \includegraphics[scale = \picscale]{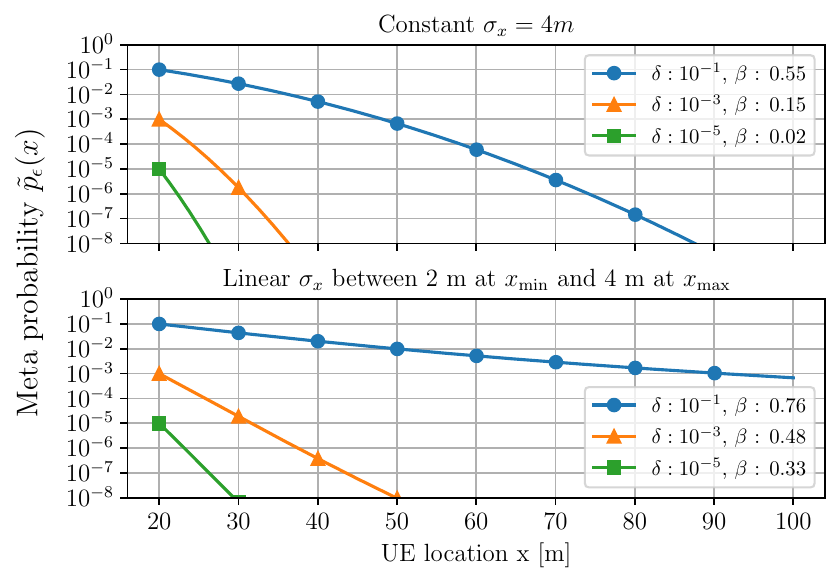}
    \caption{Meta-probability for backoff rate selection. The upper plot have constant localization standard deviation of $4$ m, and the lower plot have $\sigma_{x} = -0.025x + 4.5$ such that $\sigma_{20 \text{ m}} = 4 $ m and  $\sigma_{100 \text{ m}} = 2$ m.}
    \label{fig:meta_1d_noshadow}
\end{figure}

Figure~\ref{fig:meta_1d_noshadow} plots some examples of the meta-probability using backoff rate selection, showing that, although the target reliabilities are met at all locations, the rate is selected too conservatively except for locations close to the \gls{bs}. Hence, the global backoff solution is inherently flawed in providing an efficient solution due to its strong dependence on the \gls{ue} location. One option to improve the backoff rate-selection scheme would be to make the backoff $\beta$ dependent on the estimated location. However, we will instead suggest an alternative approach, which provides a more efficient solution while still being analytically tractable.

\subsection{Interval Rate Selection}
Instead of selecting the rate as a function of the average \gls{snr} at the estimated location, we introduce \textit{interval rate selection}, which directly considers the localization error. The idea is to select the smallest $\epsilon$-outage capacity within an interval around the estimated location, resulting in the rate selection function
\begin{align}
    R(\hat{x}, \sigma_{x}) = \min\{ C_{\epsilon}(z) \cond z \in [\hat{x} - q\sigma_{x}, \hat{x} + q\sigma_{x}] \triangleq I(\hat{x})\}, \label{eq:Interval_Rate_1D}
\end{align}
where $2q\sigma_{x}$ is the size of the interval $I(\hat{x})$. Note that the standard deviation $\sigma_{x}$ of the location estimator is assumed to be known\footnote{Assuming that the variance of an estimator is known is a common assumption in maximum likelihood estimation. In practice, the variance of the estimate $\hat{x}$ can be approximated from the \textit{observed information} function evaluated at estimated channel parameters obtained during the localization procedure (see \cite{Kay} for further details).}. In this simple scenario where fading is dominated by path loss, the minimum rate within $I(\hat{x})$ is the right interval limit such that ${R(\hat{x},\sigma_x) = C_{\epsilon}(\hat{x} + q\sigma_{x})}$, and then the meta-probability becomes 
\begin{align}
 \tilde{p}_{\epsilon}(x) &= P_{\hat{x}}(C_{\epsilon}(\hat{x} + q\sigma_{x}) > C_{\epsilon}(x)) \nonumber \\
 %&=  P_{\hat{x}}((\hat{x} + q\sigma_{x})^{-\eta} > x^{-{\eta}}) \nonumber \\
 &=  P_{\hat{x}}(\hat{x} < x - q\sigma_{x}) = \Phi(-q),\label{eq:meta_interval_1d_noshadow}
\end{align}
where \eqref{eq:rayleigh_outage} has been used. 
The scenario for interval rate selection is similar to the one illustrated in Fig.~\ref{fig:one_location_1d_no_shadow}, but instead of scaling the average \gls{snr} by a factor $\beta$, it is shifted to the left by $q\sigma_{x}$, making the distance to the outage region $x - q\sigma_{x}$. Additionally, \eqref{eq:meta_interval_1d_noshadow} shows that the dependence on location in the meta-probability is canceled out, which enables high spectral efficiency for all locations while still meeting the meta-probability requirement. Specifically, by selecting $q = -\Phi^{-1}(\delta)$, we get $\tilde{p}_{\epsilon}(x) = \delta$ for all $x$.

\subsection{Throughput}

\begin{figure}
    \centering
    \includegraphics[scale = \picscale]{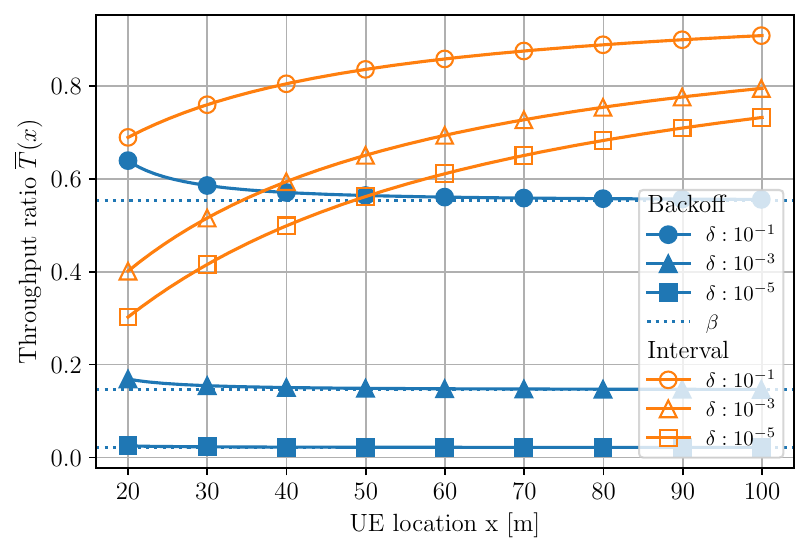}
    \caption{Throughput ratio for backoff and interval rate selection at different locations and different target confidence $\delta$. The dotted horizontal lines show the backoff value $\beta$ for $\delta = 10^{-1}, 10^{-3}, 10^{-5}$, respectively. $\sigma_{x} = 4$ m for all locations.}
    \label{fig:throughput_1d_no_shadow}
\end{figure}

The backoff and interval rate selection schemes are compared in Fig.~\ref{fig:throughput_1d_no_shadow}\footnote{The throughput in \eqref{eq:Througput_ratio} is evaluated utilizing numerical integration methods. The expectation is simplified using ${1 - p_{\text{out},x}(R(\hat{x})) = (1-\epsilon)^{\beta(x/\hat{x})^{\eta}}}$ for backoff and ${1 - p_{\text{out},x}(R(\hat{x},\sigma_x)) = (1-\epsilon)^{(x/(\hat{x} + q\sigma_x))^{\eta}}}$ for interval rate selection.} in terms of the throughput ratio $\overline{T}(x)$ from \eqref{eq:Througput_ratio}. In general, we can observe that the backoff approach is excessively conservative, with a throughput ratio depending weakly on the \gls{ue} location and converging to $\beta$ as $x$ increases.. 
For interval rate selection, we see a larger dependence on the \gls{ue} location with a throughput ratio that is significantly higher than for backoff rate selection under the same target $\delta$. The throughput ratio increases as the \gls{ue} moves away from the \gls{bs}, and numerical evaluations reveal that $\overline{T}(x) \to 1$ as $x \to \infty$ for the interval approach. 

The increase of $\overline{T}(x)$ with $x$ for the interval rate selection scheme is readily explained by the coherence radius in \eqref{eq:coherence_analytical}. For large $x$, the coherence radius is also large, and thus the localization error has less impact, i.e., $ C_{\epsilon}(\hat{x}) \approx C_{\epsilon}(x) $ with high probability resulting in $\overline{T}(x) \approx 1$. However, since the backoff approach uses a global $\beta$, even if the coherence radius is so large that the localization error is negligible, the rate is still chosen with a backoff from the average \gls{snr} leading to overprovisioning and hence low spectral efficiency.

We shall see that these observations for the simple scenario here also generalize to more complex settings. Specifically, we will now turn our attention to a more realistic wideband scenario with two-dimensional localization. 

\section{Rate Selection with Wideband Multipath Channel} \label{sec:wideband}
We now leverage the previous results to a more realistic two-dimensional scenario with the general path-based channel model in \eqref{eq:sysmod}. Albeit closed-form solutions are not possible in this case, the different rate selection functions are calibrated numerically, and the performance is illustrated through simulations in an exemplary cellular scenario. 

\subsection{Scenario and Simulated Data} \label{subsec:2Dscanario}
\begin{figure}
    \centering
    \input{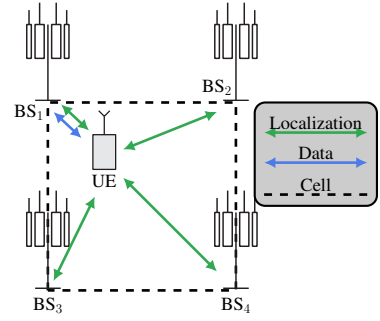}
    \caption{Exemplary scenario for two-dimensional location-based rate selection.}
    \label{fig:cell_scenario}
\end{figure}

The considered scenario is a square cell $\mathcal{R}$ of size $100 \times 100$ m$^2$ with a \gls{ue} at location $\mbf{x} \in \mathcal{R}$ and four \glspl{bs}, one in each corner, as illustrated in Fig. \ref{fig:cell_scenario}.
The \gls{ue} follows the protocol in Sec. \ref{sec:model} of first estimating its location using signals from all four \glspl{bs} and then selecting the communication rate for transmission with \gls{bs}$_1$. The channel between each \gls{bs} and the \gls{ue} follows the generic model in \eqref{eq:sysmod}, where the first path is the \gls{los} component while the remaining paths $(k > 1)$ represent the scattering. We simulate the parameters of \eqref{eq:sysmod}  --- namely $a_k$ and $\tau_k$ for $k =1,\dots,K$ --- for each location in a grid of points within the cell with grid spacing $\Delta x$\footnote{An area around the cell is also simulated to avoid edge effects when computing the meta-probability and throughput ratio --- we refer to the \href{https://github.com/TobiasKallehauge/Localization-and-Reliability-in-URLLC}{supplementary resources} for details on this.}. The parameters are simulated with the tool \textit{QuaDRiGa} following the 3GPP NR Urban Micro-Cell scenario with \gls{los} (see \cite[p. 81]{Quadriga}). The simulation parameters are summarized in Table \ref{tab:parameters}. 

\begin{table}
    \centering
    \caption{Channel and simulation parameters  } \label{tab:parameters}
    \begin{tabular}{ccc}
    \textbf{Description} & \textbf{Symbol}  & \textbf{Value} \\
    \hline & & \\[-1.5ex]
    Cell area & $\mathcal{R}$ & $[-50,50] \times [-50, 50]$ m$^2$ \\
     \gls{bs} heights & & $10$ m \\
     \gls{ue} height & & $1.5$ m \\
     Grid spacing & $\Delta x$  & $1.42$ m \\
     QuaDRiGa model & & \texttt{3GPP\_3D\_UMi\_LOS} \\
     Number of paths & $K$ & $10$ \\
     Rx and Tx antenna & & Dipole (vertical polarization) \\
     Centre frequency & $f_c$ & $3.6$ GHz \\
     Bandwidth & $W$ & 20 MHz \\
     Number of subcarriers & $N$ & 601 \\
     Transmit \gls{snr} per subcarrier & $\frac{P_{\text{tx}}}{W N_0}$ & 60 dB \\
     Target outage probability & $\epsilon$ & $10^{-3}$\\
     Confidence parameter & $\delta$ & $5\%$
    \end{tabular} 
\end{table}

\begin{figure} 
    \centering
    \includegraphics[scale = \picscale]{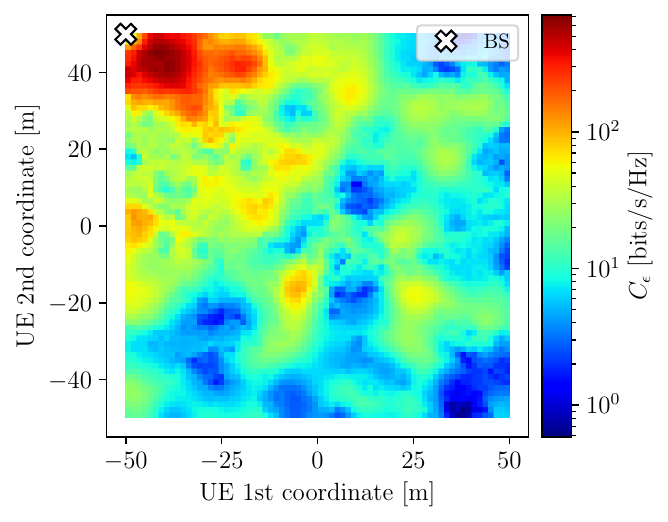}
    \caption{$\epsilon$-outage capacity when the \gls{ue} communicates with \gls{bs}$_1$ for simulation parameters in Table \ref{tab:parameters}.}
    \label{fig:R_eps_map}
\end{figure}

Aiming at reproducing the impact of fast fading, an arbitrary number of channel realizations are obtained by adding a random phase shift to each multipath component, according to the method in  \cite{Molisch2002capacity}. Hence, the $i$-th channel realization at subcarrier $j$ is obtained as
\begin{align}
h_{i,j} =\sum_{k=1}^K a_k d_j(\tau_k) e^{\jmath\theta_{i,k}} ,\quad  j = 0, \dots, N - 1,
\end{align}
where $\theta_{i,k} \overset{\text{iid.}}{\sim} \text{uniform}[-\pi,\pi)$. The ground-truth $\epsilon$-outage capacity $C_{\epsilon}$ is estimated numerically as the $\epsilon$-quantile of $10^5$ simulations of the instantaneous capacity $C$ in \eqref{eq:capacity}, and depicted in Fig.~\ref{fig:R_eps_map} --- see the \href{https://github.com/TobiasKallehauge/Localization-and-Reliability-in-URLLC}{supplementary resources} for further details on the simulations. It is clearly seen that the $\epsilon$-outage capacity is affected by path loss depending on the distance to the \gls{bs}, but local variations, commonly referred to as \textit{shadowing}, are also observed. We again see that $C_{\epsilon}$ is a smooth function of the \gls{ue} location $\mbf{x}$. 

\subsection{Localization Error}
The location is estimated using signals from the four \glspl{bs}, denoted by $\mbf{Y} =\{\mbf{y}_1,\dots,\mbf{y}_4\}$, such that $\hat{\mbf{x}} = g(\mbf{Y})$, where $g$ is some \gls{toa}-based estimator. We assume that $g$ is a consistent --- i.e., estimates converge to the true location --- maximum-likelihood estimator, which is sufficient for the asymptotic result that the estimated location is Gaussian distributed as \cite[ch. 7]{Kay}
\begin{align}
\hat{\mbf{x}} \sim \mathcal{N}(\mbf{x},\mbf{\Sigma}_{\mbf{x}}),  \label{eq:localization}
\end{align}
where the covariance matrix $\mbs{\Sigma}_{\mbf{x}}$ equals the Cramér-Rao lower bound, derived in App. \ref{sec:localization}. The appendix also provides the \gls{peb} defined as a lower bound of the \textit{root-mean-square-error} of the location estimate, i.e.,
\begin{align}
     \sqrt{E[\norm{\hat{\mbf{x}} - \mbf{x}}^2]} \geq \text{PEB}.
\end{align}
\begin{figure}
    \centering
     \includegraphics[scale = \picscale]{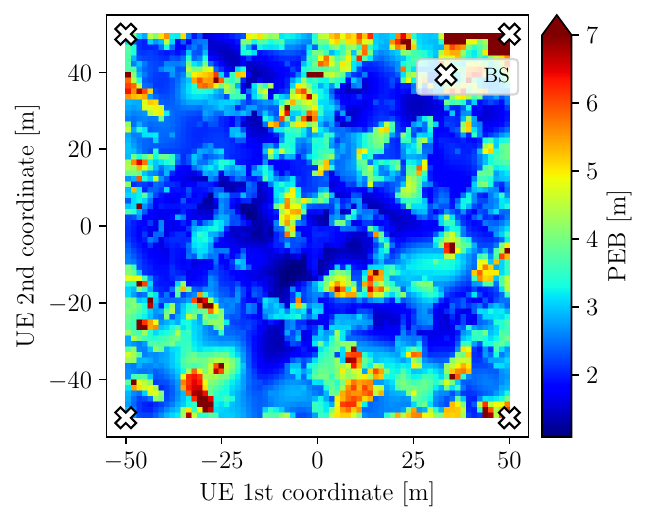}
    \caption{PEB for simulation parameters in Table \ref{tab:parameters}.}
    \label{fig:peb}
\end{figure} 
Figure~\ref{fig:peb} shows the \gls{peb} at each \gls{ue} location in our exemplary scenario.
Comparing the \gls{peb} map in Fig.~\ref{fig:peb} with the outage capacity map in Fig.~\ref{fig:R_eps_map}, no clear correlations are observed, likely because the \gls{peb} depends on the channel conditions for all four \glspl{bs}, whereas the $\epsilon$-outage capacity only depends on \gls{bs}$_1$ as illustrated in Fig. \ref{fig:cell_scenario}. Moreover, a strong multipath component is detrimental to the performance of \gls{toa} but not necessarily for communications.

To separate the effects of changing channel statistics (in Fig. \ref{fig:R_eps_map}) and localization statistics (in Fig. \ref{fig:peb}), we will also analyze a simpler form of the problem in which the covariance matrix is on the form $\mbs{\Sigma}_{\mbf{x}} = \sigma^2 \mbf{I}_{2}$ where $\sigma^2 > 0$ is constant for all locations. Sections \ref{subsec:constant_loc_error} and \ref{subsec:loc_error_different_levels} will therefore assume a constant localization variance and Sec. \ref{subsec:CRLB_loc_error} will showcase the more general scenario, where the localization variance is derived from the Cramér-Rao lower bound. Before proceeding to the results, three rate selection functions are introduced by generalizing the approaches from Sec.~\ref{sec:narrowband_rayleigh}.

\subsection{Two-dimensional Rate Selection}
\label{subsec:RateFunctions}
\subsubsection{Backoff Approach} 
\begin{figure}
    \centering
    \includegraphics[scale = \picscale]{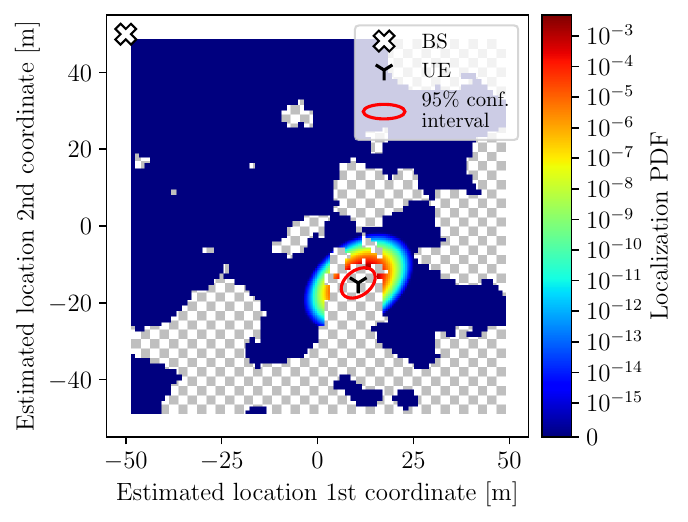}
    \caption{Outage region $S(\mbf{x})$ for simulation parameters in Table \ref{tab:parameters}, \gls{ue} location $\mbf{x} = (11, -15, 1.5)$ m, and $\mbs{\Sigma}_{\mbf{x}} = \begin{bmatrix} 3.3 & 1.1 \\ 1.1 & 2.6 \end{bmatrix}$ (computed as in App. \ref{sec:localization}). The colored and checkered area represents inside and outside the outage region (within the cell), respectively, the colors inside the outage region are the localization \gls{pdf} values, and the red ellipse represents the $95$ \% confidence interval for estimated locations.}
    \label{fig:outage_region_wideband}
\end{figure}

Since selecting the rate with a backoff from the average \gls{snr} as in \eqref{eq:rate_select_general_1d_no_shadow} is not analytically tractable here, we choose a similar approach of selecting the rate proportional to the $\epsilon$-outage capacity at the given location, i.e., 
\begin{align}
    R(\hat{\mbf{x}}) = \beta C_{\epsilon}(\hat{\mbf{x}}), \quad  \beta \in (0,1]. \label{eq:backoff_wideband}
\end{align}
The meta-probability at \gls{ue} location $\mbf{x}$ then becomes 
\begin{align}
     \tilde{p}_{\epsilon}(\mbf{x}) = P_{\hat{\mbf{x}}}(\beta C_{\epsilon}(\hat{\mbf{x}}) > C_{\epsilon}(\mbf{x}))  = P_{\hat{\mbf{x}}}(\hat{\mbf{x}} \in S(\mbf{x}))),
\end{align}
where $S(\mbf{x})$ is the outage region defined as the set of locations where the selected rate exceeds the $\epsilon$-outage capacity at location $\mbf{x}$, i.e$.$,
\begin{align}
    S(\mbf{x}) = \{\hat{\mbf{x}} \in \mathbb{R}^2 \cond \beta C_{\epsilon}(\hat{\mbf{x}}) > C_{\epsilon}(\mbf{x})  \}.
\end{align}
Figure~\ref{fig:outage_region_wideband} shows the outage region for $\beta=0.5$. We observe that, similarly to the one-dimensional case, the outage probability is \textit{generally} violated if the \gls{ue} thinks it is closer to the \gls{bs} than it actually is. However, due to local variations caused by shadowing and other propagation effects, the outage region is not perfectly circular as one may expect from the previous section, and the border of $S(\mbf{x})$ is diffuse. Interestingly, despite having a \gls{peb} of $2.4$ m in this case, which makes localization errors larger than a few meters quite unlikely, the outage probability is not-negligible, and the meta-probability is $0.48$ in this case. 

\subsubsection{Interval Approach} Generalizing the approach for the one-dimensional case, we here select the minimum value of the $\epsilon$-outage capacity within an area around the estimated location. Inspired by the confidence interval for the mean of a multivariate normal distribution,  the rate is selected as\footnote{Selecting $q^2$ as the $(1-\alpha)$-quantile of the chi-squared distribution with $2$ degrees of freedom provides a $1-\alpha$ confidence interval for the true location $\mbf{x}$.}
\begin{align}
    R(\hat{\mbf{x}}, \mbs{\Sigma}_{\mbf{x}}) &= \min\{C_{\epsilon}(\mbf{z}) \cond \mbf{z} \in I(\hat{\mbf{x}}; \mbs{\Sigma}_{\mbf{x}})\}, \label{eq:Rate_Interval}\\
    \quad I(\hat{\mbf{x}}; \mbs{\Sigma}_{\mbf{x}}) &= \{\mbf{z} \in \mR^2 \cond  (\mbf{z} - \hat{\mbf{x}})^{\T}\mbs{\Sigma}_{\mbf{x}}^{-1}(\mbf{z} - \hat{\mbf{x}}) \leq q^2\} \label{eq:interval2d}
\end{align}
where, as in \eqref{eq:Interval_Rate_1D}, $q$ is a parameter to control the size of the interval. Hence, the interval rate selection approach uses knowledge of the localization error as in Sec. \ref{sec:narrowband_rayleigh}, but in the two-dimensional case, the statistics characterized by $\mbs{\Sigma}_{\mbf{x}}$ are richer with variances along both axes and the correlation between the error in each direction. 

\subsubsection{Distance Approach} To study the importance of accounting for the local positioning error, we introduce a simplified version of the interval rate selection, referred to as the \textit{distance approach}, which chooses the minimum rate within a circle of radius $d$, i.e.,
\begin{align}
    R(\hat{\mbf{x}}) &= \min\{C_{\epsilon}(\mbf{z}) \cond \norm{\mbf{z} - \hat{\mbf{x}}}^2 \leq d^2\}. \label{eq:Rate_Distance} 
\end{align}
Note that \eqref{eq:Rate_Interval} reduces to \eqref{eq:Rate_Distance} if $\mbs{\Sigma}_{\mbf{x}}=\sigma^2\mbf{I}_2$.

\subsection{Numerical Results with Constant Localization Error} \label{subsec:constant_loc_error}

\begin{figure}[t]
    \centering
    \includegraphics[scale =.45]{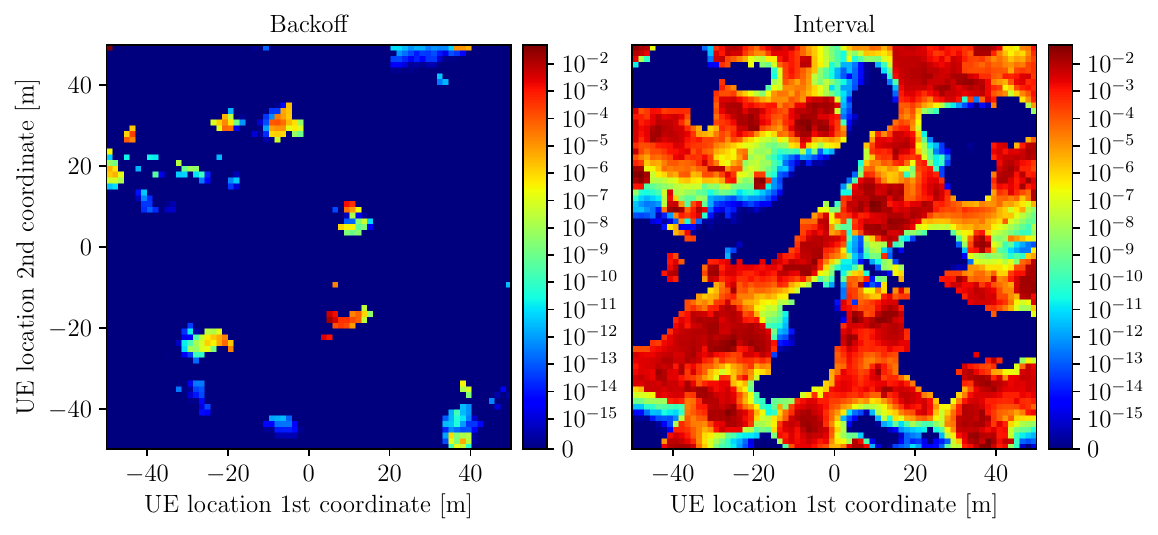}
    \caption{Meta-probability for different \gls{ue} locations with backoff and interval rate selection under constant localization error.}
    \label{fig:meta2d_const_loc_eror}
\end{figure}
\begin{figure}
    \centering
    \includegraphics[scale = .45]{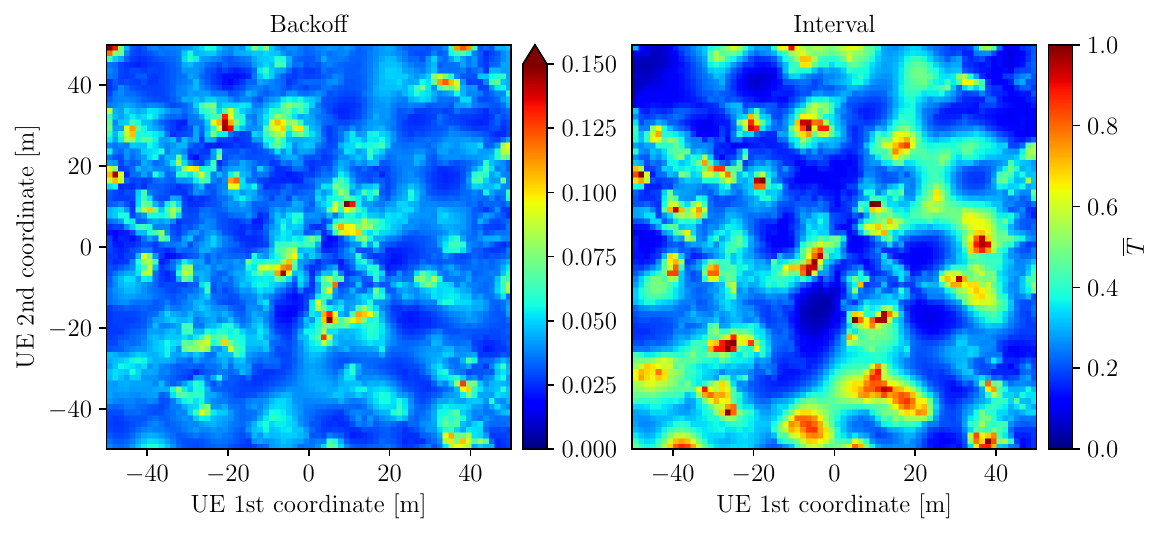}
    \caption{Throughput ratio for different \gls{ue} locations with backoff and interval rate selection under constant localization error. Note that both plots have different color mappings, as illustrated with the color bars.}
    \label{fig:throughput2d_const_loc_eror}
\end{figure}

As mentioned before, to decouple the impact of the channel conditions in both the rate selection and the localization accuracy, we first assume a fixed localization error on the form $\mbs{\Sigma}_{\mbf{x}} = \sigma^2 \mbf{I}_{2}$ and where $\sigma^2 > 0$ is constant for all locations. Under these conditions, we evaluate the performance of the different rate selection functions in our exemplary scenario (simulation parameters in Table \ref{tab:parameters}). Note that, since $\mbs{\Sigma}_{\mbf{x}} = \sigma^2 \mbf{I}_{2}$, the distance rate selection is equivalent to the interval approach; hence, the former is neglected here. 

\begin{figure}[t]
    \centering
    \includegraphics[scale = \picscale]{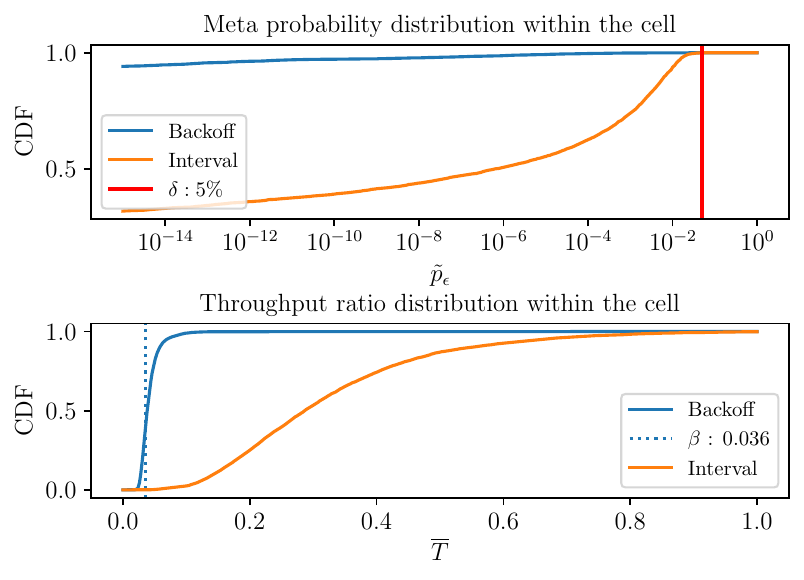}
    \caption{Distribution of meta-probability (top) and throughput ratio (bottom) for all \gls{ue} locations within the cell.}% The average throughput for the backoff and interval approach are $0.042$ and $0.321$, respectively.}
    \label{fig:cdf_constant_loc}
\end{figure}

We fix $\sigma^2 = 12.5$ m$^2$ --- which gives a \gls{peb} of $5$ m --- and solve \eqref{eq:opmization} numerically for both the backoff and interval approaches, yielding $\beta = 0.036$ and $q^2 = 5.92$ for a confidence $\delta = 5$ percent. The resulting meta-probability and throughput ratio are depicted in Figs. \ref{fig:meta2d_const_loc_eror}-\ref{fig:throughput2d_const_loc_eror}, respectively, and the distribution of both performance indicators across the cell is plotted in Fig. \ref{fig:cdf_constant_loc}. 
 
As in Sec. \ref{sec:narrowband_rayleigh}, we notice an excessively conservative rate for the backoff approach, rendering a meta-probability much lower than required in a large portion of the cell. While not reaching $\delta$ at all locations, as in the Rayleigh case, the interval approach is generally closer to the target meta-probability. This is better observed in Fig. \ref{fig:cdf_constant_loc}, where it is also seen that the throughput ratio obtained by the backoff approach is considerably lower than the interval approach --- specifically, an average throughput ratio of $0.042$ and $0.321$ for the backoff and interval approaches, respectively. Again, the global selection of $\beta$ is penalizing the spectral efficiency of the system.

Compared to the simple scenario in Sec. ~\ref{sec:narrowband_rayleigh}, a key difference is that the distance to the \gls{bs} is no longer a good performance indicator --- in the one-dimensional Rayleigh case, improved performance was obtained as the \gls{ue} moved farther away from the \gls{bs} due to increased coherence radius. In this more realistic case, local variations of the $\epsilon$-outage capacity due to different shadowing and fading conditions must be considered (see Fig.~\ref{fig:R_eps_map}).

\begin{figure*}[t]
    \centering
    \includegraphics[scale = \picscale]{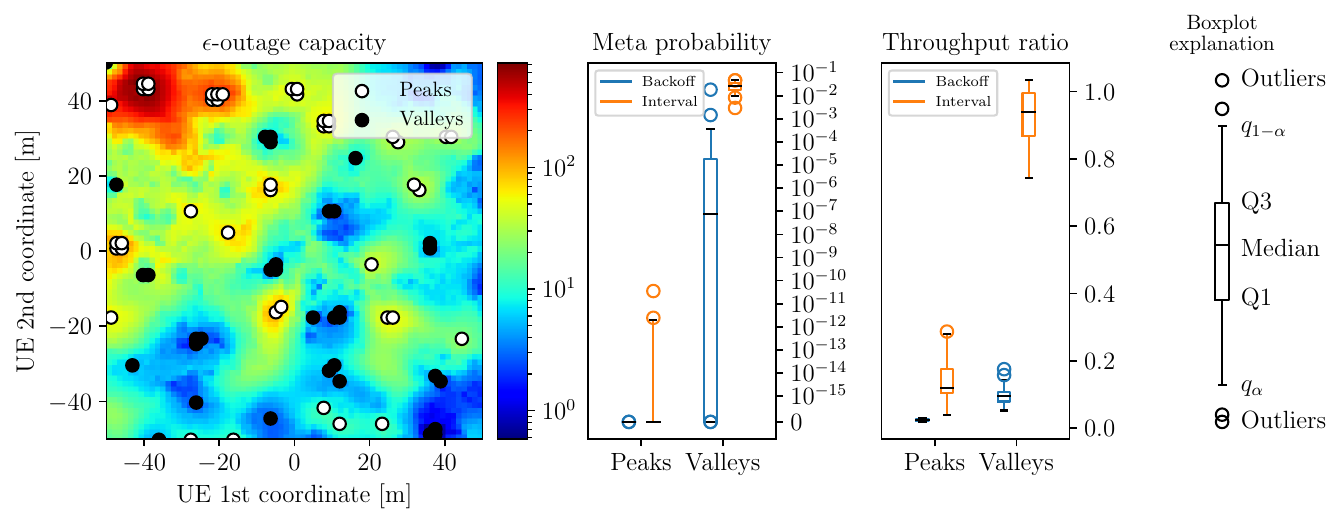}
    \caption{Left: Location of local maxima (peaks) and local minima (valleys) of the $\epsilon$-outage capacity. Right: Statistics for the meta-probability and throughput ratio as boxplots grouped by either peaks or valleys.}
    \label{fig:peak_valey}
\end{figure*}

To further study the relation between the $\epsilon$-outage capacity and performance (i.e., how Fig. \ref{fig:R_eps_map} relates to Figs. \ref{fig:meta2d_const_loc_eror} and \ref{fig:throughput2d_const_loc_eror}), we observe that low values of meta-probability and throughput ratio are correlated with local maxima of $C_{\epsilon}$ whereas, in turn, high values are associated with local minima of $C_{\epsilon}$. This relation is quantified in Fig. \ref{fig:peak_valey}, where local maxima (peaks) and minima (valleys) of the $\epsilon$-outage capacity are detected\footnote{The peaks and valleys are detected manually based on visual inspection.}, and the results at the detected locations are then grouped by either peak or valley. In this example, $41$ peaks and $32$ valleys are detected within the cell, and the statistics for the meta-probability and throughput ratio are summarized by boxplots, showing the median, the $1$st and $3$rd quartile, and the $\alpha$ and $1-\alpha$ quantiles of the data ($\alpha=5\%$).

The figure confirms the previous observation, showing that the meta-probability for the interval approach is only close to the target value $\delta$ at the valleys, and the throughput ratio is also only close to $1$ at the valleys. The same relation holds for the backoff approach, however, the overall poorer performance makes the difference less noticeable. To understand the effect of peaks and valleys, consider first that the \gls{ue} is located at a peak of the $\epsilon$-outage capacity. Since $C_\epsilon(\mbf{x})$ is, by definition, larger at the peak than the surrounding locations, any error in the location estimation will make the \gls{ue} to believe its channel conditions are worse than the actual ones, leading to conservative rate selection and low meta-probability and throughput. Conversely, at valley locations, the \gls{ue} almost surely thinks it has better channel conditions making rate selection less conservative, yielding larger meta-probabilities and throughput ratios. Hence, we see here that local minimum of $C_\epsilon(\mbf{x})$ have the highest probabilities of violating the outage requirement while simultaneously having high throughput ratio.

\begin{figure}
    \centering
    \includegraphics[scale = \picscale]{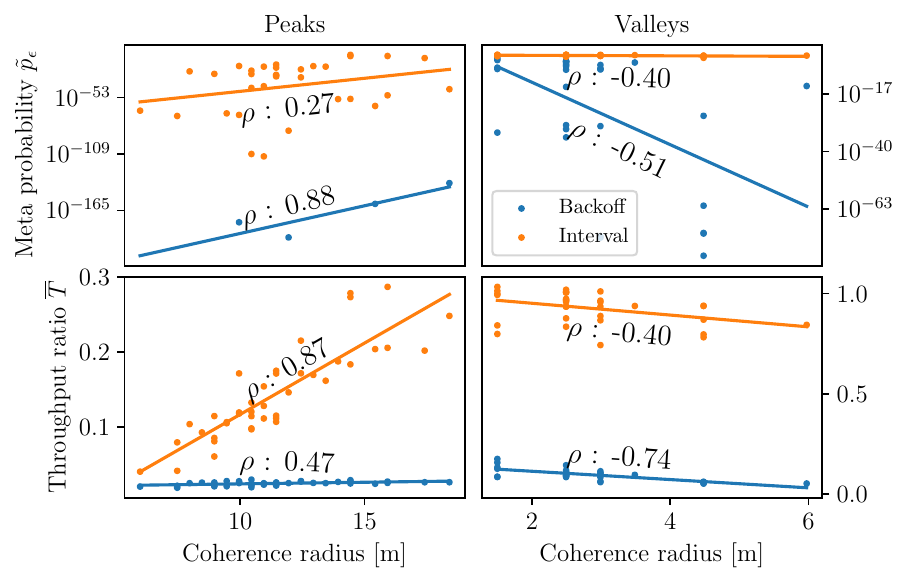}
    \caption{Meta probability and throughput ratio at peak and valley locations plotted against the coherence radius using backoff and interval rate selection. The fitted best straight lines and the correlation coefficients $\rho$ are shown.}
    \label{fig:coherence_dependencies}
\end{figure}

Aiming at getting more insight, we re-introduce the notion of coherence radius as a generalization of \eqref{eq:coherence_1d} for two dimensions as 
\begin{align}
    \text{CR}(\mbf{x}) = \min_{\varrho > 0}\left\{ \varrho \hspace*{1pt} \left|\hspace*{1pt} \max_{\mbf{z} \in B_{\varrho}(\mbf{x})} \left\{\frac{|C_{\epsilon}(\mbf{x}) - C_{\epsilon}(\mbf{z})|}{C_{\epsilon}(\mbf{x})} \right\} > t \right. \right\},
\end{align}
where $B_{\varrho}(\mbf{x})$ is a disk of radius $\varrho$ centered at location $\mbf{x}$. Using $t = 0.9$,
the relation of coherence radius with meta-probability and throughput ratio is illustrated in Fig \ref{fig:coherence_dependencies}, suggesting a noticeable correlation most pronounced for the throughput ratio with correlations in the range $[0.40,0.87]$.
Interestingly, the coherence radius at the peaks correlates positively with the meta-probability and throughput ratio and vice-versa for the valleys. Hence, the coherence radius provides a continuous measure of the impact of the \gls{ue} being in a peak or valley, e.g., a peak with a low coherence radius tends to have lower throughput ratio than a peak with a higher coherence radius. However, we do not see a one-to-one relation as in Sec. \ref{sec:narrowband_rayleigh}, so while the coherence radius certainly affects the meta-probability and throughput ratio, it is too simplistic to fully characterize the performance of location-based rate selection under realistic propagation conditions (e.g., due to the local variations of $C_{\epsilon}$ caused by shadowing). 

Albeit the peak/valley classification explains the system performance at the two extreme cases, the $\epsilon$-outage capacity at the rest of the locations may have a different trend depending on the direction (as shown in Fig. \ref{fig:R_eps_map}), i.e., the sign of the gradient of $C_\epsilon(\mbf{x})$ depends on the radial direction. Therefore, it is hard to extract a general conclusion unless full knowledge of the two-dimensional spatial behavior of $C_\epsilon(\mbf{x})$ is available. Nonetheless, for a specific spatial direction, the previous findings  still hold. 

\subsection{Numerical Results with Different Levels of Constant Localization Error} \label{subsec:loc_error_different_levels}
\begin{figure}[t]
    \centering
    \includegraphics[scale = \picscale]{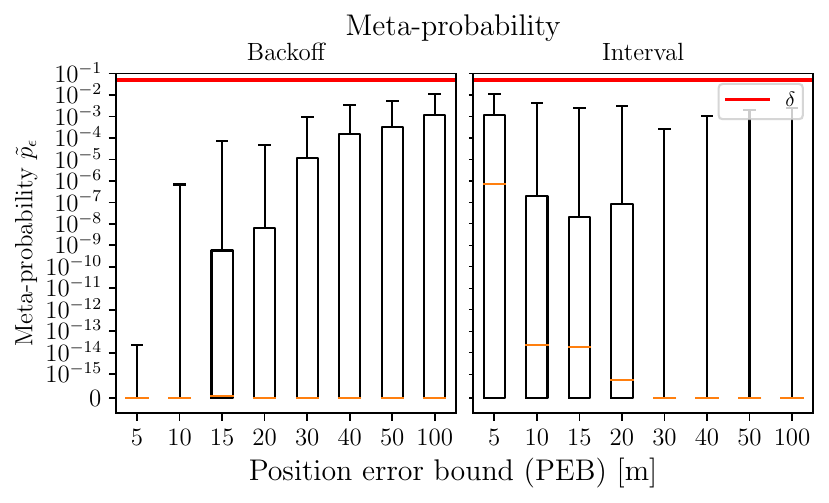}
    \includegraphics[scale = \picscale]{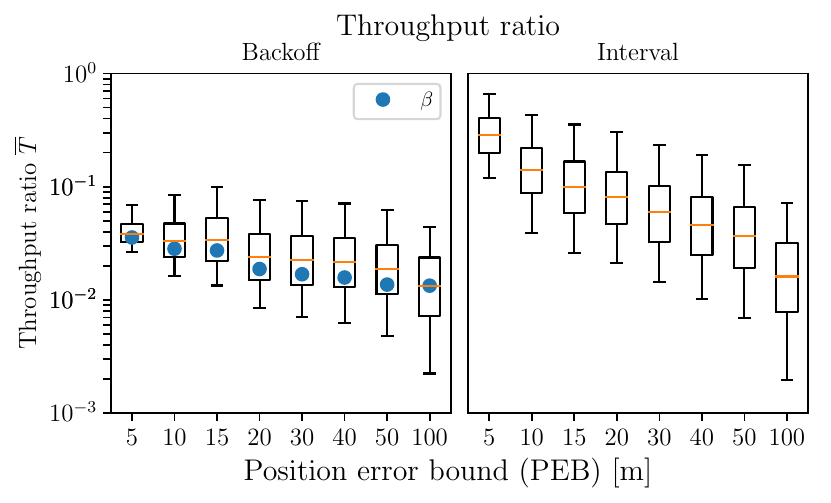}
    \caption{Distribution of meta-probability (top) and throughput ratio (bottom) for backoff and interval approach within the cell for different \glspl{peb} illustrated as boxplots. Note that the throughput ratio is plotted on a logarithmic scale. The chosen backoff $\beta$ for the backoff approach is shown in the plot second to the right. Outliers outside the $q_{\alpha}$ and $q_{1 - \alpha}$ quantile with $\alpha = 5\%$  are not shown.}
    \label{fig:different_peb}
\end{figure}

We now evaluate the impact of the localization error on the system performance; that is, whether the previous conclusions can be generalized for different localization errors. To that end, we assume again $\mbs{\Sigma}_{\mbf{x}} = \sigma^2 \mbf{I}_{2}$ and vary the value of $\sigma^2$. Specifically, Fig. \ref{fig:different_peb} shows results for $\sigma^2$ between $12.5$ m$^2$ and 5000 m$^2$, corresponding to a \gls{peb} between $5$ m and $100$ m. Note that the latter represents a quite large location error, but we intentionally included it to see the system performance in extreme cases. 

Observing Fig. \ref{fig:different_peb}, we notice that, for the backoff approach, the meta-probability gets closer to the target value $\delta$ as the localization error variance increases. To understand this, recall that the localization error does not change the shape of the outage region (see Fig. \ref{fig:outage_region_wideband}) but increases the probability of estimating a location within that region (i.e., higher meta-probability). Conversely, the meta-probability decreases with larger location errors for the interval rate selection since it increases the area of the interval, and hence finding a local minimum of $C_\epsilon(\mbf{x})$ is more likely. In the extreme case, the interval accounts for the whole cell; thus, the global minimum is selected. The impact on the throughput ratio is the same for both rate selection approaches, with decreasing throughput ratio for larger localization variance. However, the relative impact on the performance is larger for the interval approach.

\begin{figure}
    \centering
    \includegraphics[scale = \picscale]{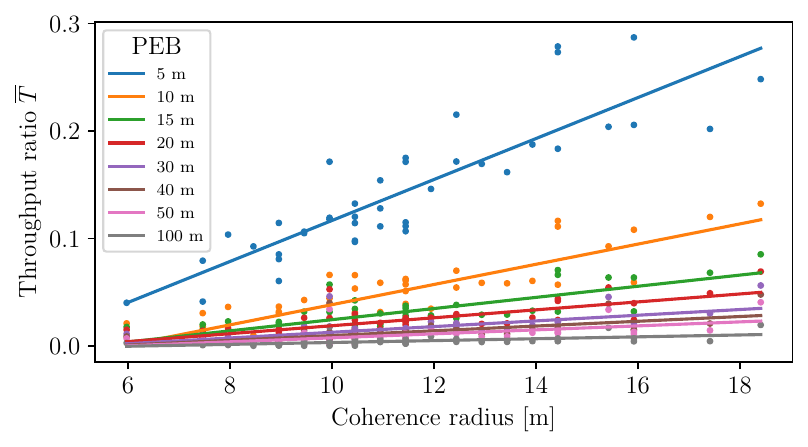}
    \caption{Throughput ratio at the peak locations in Fig. \ref{fig:peak_valey} for the interval approach at different \glspl{peb}.}
    \label{fig:thorughput_different_peb}
\end{figure}

Another result is how the localization variance affects the relationship between coherence radius and performance. It is found that increasing the localization variance has the effect of lessening the dependence on the coherence radius  (i.e., decreasing the correlation). This is because larger location errors diminish the influence of local variations in $\epsilon$-outage capacity and hence how the coherence radius affects the performance. We see this trend in Fig. \ref{fig:thorughput_different_peb}.

\subsection{Numerical Results with Error Bound-Based Localization Error} \label{subsec:CRLB_loc_error}

\begin{figure*}
    \centering
    \includegraphics[scale = \picscale]{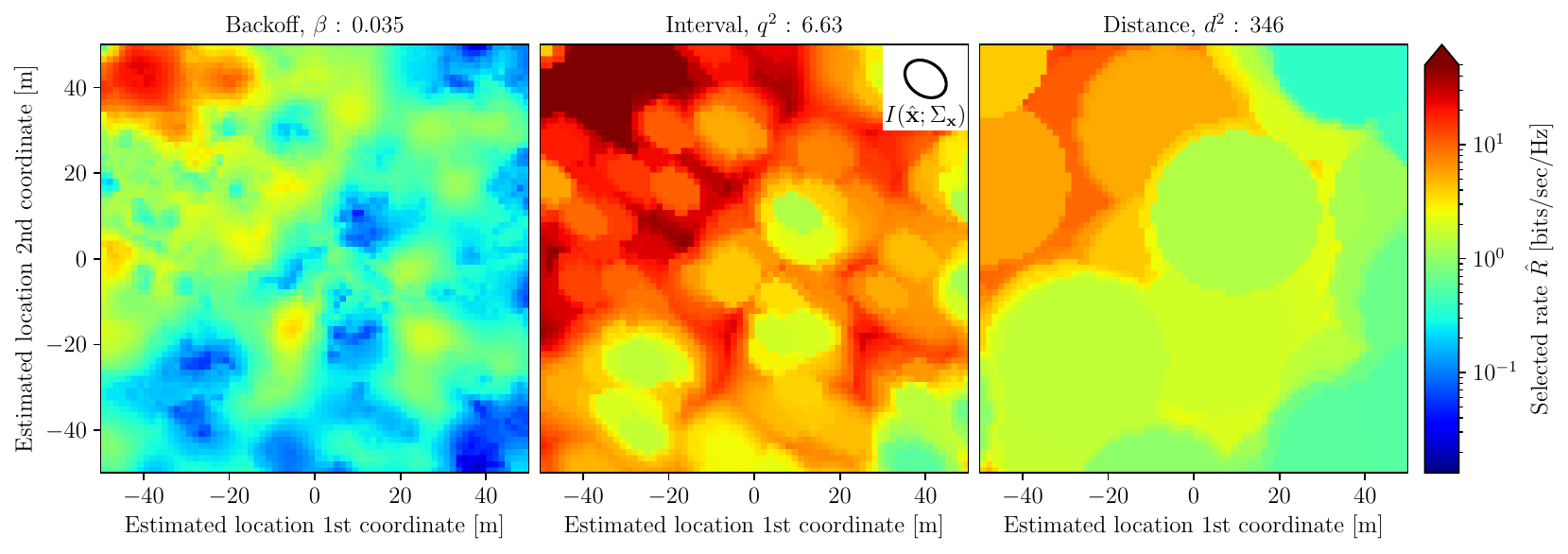}
    \caption{Selected rate with the backoff, interval, and distance approach for different estimated locations with simulation parameters in Table \ref{tab:parameters}.}
    \label{fig:rate_2d}
\end{figure*}

As a final result, we now evaluate the system's performance in the full scenario wherein the localization error is based on the error bounds derived in App. \ref{sec:localization} and illustrated in Fig. \ref{fig:peb}, i.e., the localization error covariance matrix is now dependent on the location and not a diagonal matrix. As previously, \eqref{eq:opmization} is solved numerically for the three rate selection functions in Sec. \ref{subsec:RateFunctions}, yielding $\beta = 0.035$ for backoff, $q^2 = 6.63$ for interval, and $d^2 = 346$ for the distance approach. The resulting rates $R(\hat{\mbf{x}})$ at every estimated location $\hat{\mbf{x}} \in \mathcal{R}$ within the cell are plotted in Fig. \ref{fig:rate_2d}. Note that the rate selection approach from \eqref{eq:Rate_Interval} also depends on the localization covariance $\mbs{\Sigma}_{\mbf{x}}$, hence, $R(\hat{\mbf{x}},\mbf{\Sigma}_{\mbf{x}})$. The rates shown in the center plot all correspond to $\mbs{\Sigma}_{\mbf{x}}$ at  true \gls{ue} location $\mbf{x} = (-26, -16,   1.5)$ m, which defines the shape of the interval $I(\hat{\mbf{x}}; \mbs{\Sigma}_{\mbf{x}})$ illustrated in the top right corner.

Fig.~\ref{fig:rate_2d} immediately reveals that the selected rates in the interval approach are considerably larger than the two other approaches, potentially leading to higher spectral efficiency. While the chosen rates with the backoff approach are simply a scaled version of the $\epsilon$-outage capacity, the interval and distance approach shows different spatial patterns, both with several areas where the selected rate is flat due to a local minimum in the capacity at the center of the flat areas. 

\begin{figure}
    \centering
    \includegraphics[scale = \picscale]{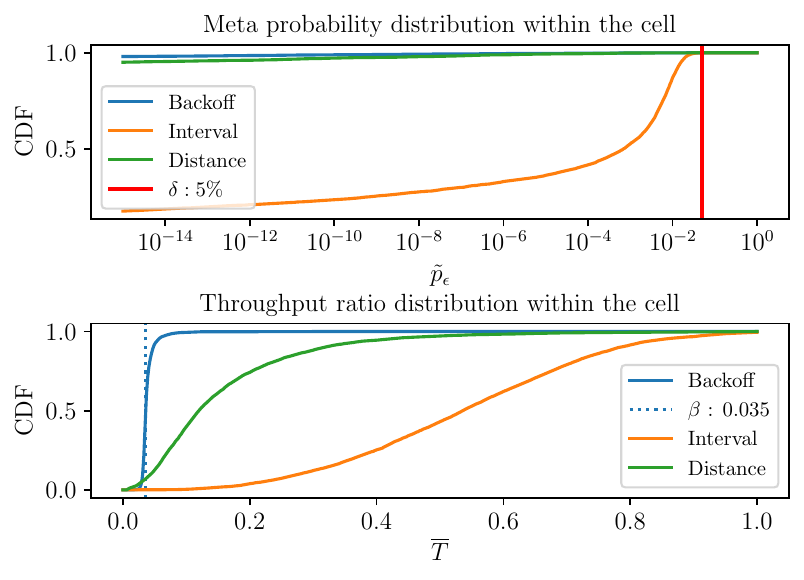}
    \caption{Distribution of meta-probability (left) and throughput ratio (left) for all \gls{ue} locations within the cell with parameters $\beta = 0.035$ (backoff), $q^2 = 6.63$ (interval), and $d^2 = 346$ (distance).}
    \label{fig:meta_throughput_2d}
\end{figure}

Looking at the overall performance, the statistics for the meta-probability and throughput ratio are summarized in Fig. \ref{fig:meta_throughput_2d}.
We see that the results for the backoff and interval approaches are similar to those for constant localization error in Fig. \ref{fig:cdf_constant_loc}, with overall better performance for the interval approach. Similar to the results in Sec. \ref{sec:narrowband_rayleigh}, the meta-probability for the backoff approach is generally much lower than required, being hence a too conservative approach. The same holds for the distance approach, although with a slight increase in the average throughput ratio. Interestingly, despite following a similar rationale, we see that the interval rate selection approach is more efficient than the distance-based approach, highlighting the importance of accounting for the specific localization accuracy. With the added complexity in the scenario of non-constant localization error, the difference between results in peaks/valleys of the $\epsilon$-outage capacity is less pronounced but still significant (albeit not shown in the figures). The impact of the coherence radius is also lessened with correlations in the order of $|\rho| \approx 0.25$. This again points out the complexity of the location-based rate selection problem and the fact that predicting the reliability and throughput requires extensive knowledge of the scenario (i.e., simply knowing the coherence radius is not sufficient).

\section{Conclusions} \label{sec:conclusion}

Within the context of ultra-reliable communications, this paper has analyzed the impact of location uncertainty on communication reliability when the transmission rate is selected based on localization. Different cases have been studied through a rigorous statistical framework, ranging from a one-dimensional scenario subject to Rayleigh fading to a two-dimensional case with \gls{3gpp} channels and localization error derived from Fisher analysis. By isolating the location error as the only source of uncertainty, we have shown that it considerably impacts the communication reliability --- characterized by the meta-probability --- and the achievable throughput. Overall, spatial consistency of the channel statistics is pointed out as beneficial for performance, resulting in low meta-probability and high throughput ratio. However, as the propagation environment becomes more intricate (equivalently, more realistic), this correlation is lessened, and some exceptions arise. It was seen explicitly that higher coherence radius at local minima of the $\epsilon$-outage capacity causes increases in meta-probability (i.e., lower reliability) but also higher throughput ratio, and vice versa for local maxima. A comprehensive understanding of the system is, therefore, necessary to fully account for the impact of localization error, but simple measures such as the coherence radius are still informative. As expected, larger localization errors translate into a reduced performance but also reduce the dependency on variations of local channel conditions. Different rate selection functions have also been tested, highlighting that correctly accounting for localization uncertainty at each position within is of capital importance to reliably select the transmission rate without penalizing the spectral efficiency. Overall, global rate selection policies, i.e., not accounting for specific local propagation conditions, are shown to be excessively conservative.

Future work in studying the impact of localization error on ultra-reliable communication includes exploring more realistic channel models, e.g., including blockages to introduce non-smooth channel statistics or working with real datasets. Another important future direction is to include the error introduced in the process of mapping channel statistics based on location. That is, analyzing radio mapping approaches (e.g., in \cite{kallehaugeGlobecom2022}) and using the insight gained in this work to account for localization errors. This work helps to lay the foundation for understanding how localization errors impact ultra-reliable communication. We specifically studied physical layer reliability for location-based rate selection, but the results here can be generalized with the ultimate objective of enabling ultra-high reliability using localization and sensing in future networks such as \gls{6g}.

\appendix
\subsection{Error-Bounds for Time of Arrival-Based Localization} \label{sec:localization}
In this appendix, the error of \gls{toa}-based localization is modeled in the scenario described in Sec. \ref{subsec:2Dscanario} with signals from the channel in \eqref{eq:sysmod}, where the first path ($k = 1$) is the \gls{los} path and the remaining paths ($k > 1)$ are \gls{nlos} scattering. \Gls{toa} localization is assumed, based on the propagation delay of the \gls{los} path and requiring $D+1$ reference signals from different \glspl{bs}. The scenario in Fig. \ref{fig:cell_scenario} is considered, and the incoming signal from each \gls{bs} is denoted by $\mbf{y}_i$ for $i = 1,\dots,4$. \Gls{tdoa} is often used to cancel the clock bias \cite{survey_TOA}; however, we here consider \gls{toa} and introduce the effect of clock bias $B$ in the localization uncertainty. Hence, the measured \gls{los} delay for \gls{bs} $i$ at location $\mbf{x}_{\text{bs},i}$ is then $\tilde{\tau}_{i, 1} = \norm{\mbf{x} -\mbf{x}_{\text{bs},i}}/c + B$, where $c$ is the speed of light and $B\in\mathbb{R}$ is the clock bias. The variance of any unbiased estimator $\hat{\mbf{x}}$ of the true position $\mbf{x}$ is then characterized through the \gls{crlb} as
\begin{align}
    \Var[\hat{\mbf{x}}(\mbf{y}_1, \mbf{y}_{2}, \mbf{y}_{3}, \mbf{y}_4)] \geq J^{-1}(\mbf{x}),
\end{align}
where $J(\mbf{x})$ is the \gls{fim} corresponding to the \gls{ue} location $\mbf{x}$ \cite{Kay}. The first step in finding the \gls{fim} for $\mbf{x}$ is to compute the \gls{fim} with respect to \textit{all unknown parameters} in \eqref{eq:sysmod}, denoted 
\begin{equation}
        \resizebox{1\hsize}{!}{$\mbs{\eta}_i = \left[ \tilde{\tau}_{i,1} \ \dots \ \tilde{\tau}_{i,K_i'} \ \mathfrak{R}(a_{i,1}) \ \dots \ \mathfrak{R}(a_{i,K_i'}) \ \mathfrak{I}(a_{i,1}) \ \dots \ \mathfrak{I}(a_{i,K_i'}) \right]^{\T}$}
\end{equation}
for $i = 1,2,3,4$, where the first subscript in the parameters denotes the \gls{bs}, and the second denotes the path index. $K_i'$ is the number of non-resolvable paths from the $i$th \gls{bs} plus the \gls{los} path. The non-resolvable paths are defined as all \gls{nlos} scattered paths % with delay difference to the \gls{los} less than the inverse of the bandwidth, i.e., $K_i'$ is 
such that $|\tilde{\tau}_{i,1}-\tilde{\tau}_{i,K_i'}| \leq \frac{1}{W}$ and $|\tilde{\tau}_{i,1}-\tilde{\tau}_{i,K_i'+1}| > \frac{1}{W}$, where $W$ is the system bandwidth \cite{Aditya2018Wideband}. Without any loss of generality, we will assume that the number of subcarriers $N$ is odd such that $N = 2p + 1$  for $p \in \mathbb{N}$ and that the subcarriers are indexed as $j = -p,\dots, p$. Denoting $\mbf{d}(\tilde{\tau}_{i,k})$ as the vector with elements $d_j(\tilde{\tau}_{i,k}) = \exp(-2\pi\jmath \Delta_f j \tilde{\tau}_{i,k})$, the normalized received signal $\mbf{y}_i/\sqrt{P_{\text{tx}}}$ in \eqref{eq:sysmod} follows a circularly symmetric, complex Gaussian distribution with mean $\mbs{\mu}_i = \sum_{k=1}^K a_{i,k} \mbf{d}(\tilde{\tau}_{i,k})$ and covariance $\frac{W N_0}{P_{\text{tx}}}\mbf{I}_{N}$. Therefore \cite{Kay}:
\begin{align} \label{eq:fisher_eta}
J(\mbs{\eta}_i) &= \frac{2P_{\text{tx}}}{W N_0} \sum_{j=-p}^{p} \mathfrak{R}\left( \frac{\partial \mu_{i,j} }{\partial \mbs{\eta}_i}\left(\frac{\partial \mu_{i,j} }{\partial \mbs{\eta}_i}\right)^{\H}\right).
\end{align}
The partial derivatives in \eqref{eq:fisher_eta} with respect to the unknown channel parameters are
\begin{align}
    \frac{\partial \mu_{i,j}}{\partial \tilde{\tau}_{i, k}} &= -a_{i, k} \jmath 2\pi j \Delta_f \exp(-\jmath2\pi j \Delta_f \Tilde{\tau}_{i, k}),\\
    \frac{\partial \mu_{i,j}}{\partial \mathfrak{R}(a_{i,k})} &= \exp(-\jmath2\pi j \Delta_f \Tilde{\tau}_{i, k}),\\
    \frac{\partial \mu_{i,j}}{\partial \mathfrak{I}(a_{i,k})} &= \jmath\exp(-\jmath2\pi j \Delta_f \Tilde{\tau}_{i, k}).
\end{align}
Denoting by $[\mbf{A}]_{i:j,k:p}$ the submatrix of $\mbf{A}$ from row $i$ to $j$ and column $k$ to $p$, and by $\text{diag}(\mbf{a})$ the diagonal matrix formed from vector $\mbf{a}$, the \gls{fim} is expressed as
\begin{equation}
    J(\mbs{\eta}_i) = \frac{8\pi P_{\text{tx}}}{W N_0}
    \begin{pmatrix}
        \mbs{J}_{\mbs{\tau}_{i}} & \mbs{J}_{\mbs{\tau}_{i}\mbs{a}_{i}}^T \\
        \mbs{J}_{\mbs{\tau}_{i}\mbs{a}_{i}} & \mbs{J}_{\mbs{a}_{i}\mbs{a}_{i}}
    \end{pmatrix},
\end{equation}
where the involved terms are calculated as
\begin{align}
    &[\mbs{J}_{\mbs{\tau}_{i}}]_{n,m} = 4\pi \Delta_f
    \begin{cases}
        \sum_{j=1}^{N/2} j^2 \mathfrak{R}(d_j(\tau_{i,n}-\tau_{i,m}))& \\
        \times\mathfrak{R}(a_{i,n} a_{i,m}^*) , & n \neq m,\\
        |a_n|^2 \frac{N(N+1)(N+2)}{24}, & n = m,
    \end{cases}\\
    &\mbs{J}_{\mbs{\tau}_{i}\mbs{a}_{i}} = -
    \begin{pmatrix}
        \mbs{C}~\text{diag}\Big(\begin{bmatrix}\mathfrak{R}(a_{i,1}) & \cdots & \mathfrak{R}(a_{i,K_i'})\end{bmatrix}\Big)\\
        \mbs{C}~\text{diag}\Big(\begin{bmatrix}\mathfrak{I}(a_{i,1}) & \cdots & \mathfrak{I}(a_{i,K_i'})\end{bmatrix}\Big)
    \end{pmatrix},\\
    &[\mbs{C}]_{n, m} =
    \begin{cases}
        \sum_{j=1}^{N/2} j \mathfrak{I}(d_j(\tau_{i,n}-\tau_{i, m})), & n \neq m,\\
        0, & n = m,
    \end{cases}\\
    &\mbs{J}_{\mbs{a}_{i}\mbs{a}_{i}} =
    \begin{pmatrix}
        \mbs{J}_{\mathfrak{R}(\mbs{a}_{i})\mathfrak{R}(\mbs{a}_{i})} & \mbs{0}\\
        \mbs{0} & \mbs{J}_{\mathfrak{I}(\mbs{a}_{i})\mathfrak{I}(\mbs{a}_{i})}
    \end{pmatrix},\\
    &[\mbs{J}_{\mathfrak{R}(\mbs{a}_{i})\mathfrak{R}(\mbs{a}_{i})}]_{n, m} = [\mbs{J}_{\mathfrak{I}(\mbs{a}_{i})\mathfrak{I}(\mbs{a}_{i})}]_{n,m}\\
    &\quad\quad=
    \begin{cases}
        1 + 2 \sum_{j=1}^{N/2}\mathfrak{R}(d_j(\tau_{i, n} - \tau_{i, m})), & n \neq m,\\
        N + 1, & n = m.
    \end{cases}
\end{align}

In $\mbs{\eta}_i$, only the LoS delay $\tilde{\tau}_{i,1}$ contains information about the location $\mbf{x}$, so we continue with the equivalent \gls{fi}~\cite{equivalent_fisher} 
\begin{align}
    J^E(\tilde{\tau}_{i,1}) =& [J(\mbs{\eta}_i)]_{1,1}
    - [J(\mbs{\eta}_i)]_{1,2:3K_i'}([J(\mbs{\eta}_i)]_{2:3K_i', 2:3K_i'})^{-1} \notag \\
    &\times[J(\mbs{\eta}_i)]_{2:3K_i',1}
\end{align}
where the second term is interpreted as the information loss from the unknown \textit{nuisance parameters}.

Assuming independence of the signals,  $J^E(\tilde{\tau}_{1,1},\tilde{\tau}_{2,1},\tilde{\tau}_{3,1},\tilde{\tau}_{4,1})$ is the diagonal matrix with elements $J^E(\tilde{\tau}_{1,1}),J^E(\tilde{\tau}_{2,1}),J^E(\tilde{\tau}_{3,1}),J^E(\tilde{\tau}_{4,1})$, and the \gls{fim} with respect to the location $\mbf{x} = \begin{bmatrix} x_1 & x_2\end{bmatrix}^{\T}$ and clock bias $B$ is obtained using the transformation \cite{Kay}
\begin{align}
    J(\mbf{x},B) &= \mbf{T}^{\T} J^E(\tilde{\tau}_{1,1},\tilde{\tau}_{2,1},\tilde{\tau}_{3,1},\tilde{\tau}_{4,1}) \mbf{T}, \\
     \mbf{T} &= \begin{bmatrix} 
    \frac{\partial \tilde{\tau}_{1,1}}{\partial x_1} & \frac{\partial \tilde{\tau}_{1,1}}{\partial x_2} & \frac{\partial \tilde{\tau}_{1,1}}{\partial B}  \\
    \vdots & \vdots & \vdots \\
    \frac{\partial \tilde{\tau}_{4,1}}{\partial x_1} & \frac{\partial \tilde{\tau}_{4,1}}{\partial x_2} & \frac{\partial \tilde{\tau}_{4,1}}{\partial B} 
    \end{bmatrix} \\
    \frac{\partial \tilde{\tau}_{i,1}}{\partial x_d} &= \frac{x_d - (x_{\text{bs},i})_d}{c\norm{\mbf{x} - \mbf{x}_{\text{bs},i}}} \text{ for } d = 1,2 \text{ and } i = 1,2,3,4 \\
    \frac{\partial \tilde{\tau}_{i,1}}{\partial B}  &= 1, \text{ for } i = 1,2,3,4.
\end{align}
$J^{-1}(\mbf{x}) = \left[J(\mbf{x},B)^{-1}\right]_{1:2,1:2}$ now gives the \gls{crlb} as a positive definite $2\times 2$ matrix whose diagonal elements contain the variance of each coordinate, and off-diagonal elements are the covariances. This bound was derived for a specific set of parameters $\mbs{\eta}_1,\mbs{\eta}_2,\mbs{\eta}_{3}, \mbs{\eta}_4$, wherein the channel coefficients $a_{i,k}$ are modeled as random variables while the delays $\tilde{\tau}_{i,k}$ are assumed deterministic. Assuming that the channel changes quickly (e.g., in a block-fading channel), the localization model should reflect the average behavior rather than a specific realization. For this, denote $\mathcal{A} = \{\mathfrak{R}(a_{i,k}), \mathfrak{I}(a_{i,k}) : k=1,\dots,K_i',~i=1,\dots,4\}$ as the total set of all random parameters such that $J^{-1}(\mbf{x};\mathcal{A})$ is the conditional \gls{crlb} given a specific channel realization \cite{Noam2009Hybrid}. This, in turn, means that $\Var[\hat{\mbf{x}}\cond \mathcal{A}] \geq J^{-1}(\mbf{x};\mathcal{A})$, and the law of total variance then tells us that 
\begin{align}
    \Var[\hat{\mbf{x}}] &= E[\Var[\hat{\mbf{x}}\cond \mathcal{A}]] + \Var[E[\hat{\mbf{x}}\cond \mathcal{A}]] \nonumber \\
    &\overset{(a)}{=}  E[\Var[\hat{\mbf{x}}\cond \mathcal{A}]] + \underbrace{\Var[\mbf{x}]}_{=\mbf{0}} \nonumber \\
    &= \int \Var[\hat{\mbf{x}}\cond \mbf{A}]f(\mathcal{A}) \ d\mathcal{A} \nonumber \\
    &\overset{(b)}{\geq} \int J^{-1}(\mbf{x};\mathcal{A}) f(\mathcal{A}) \ d\mathcal{A} \triangleq \bar{J}^{-1}(\mbf{x}), \label{eq:avg_var}
\end{align}
where $f(\mathcal{A})$ denotes the joint distribution of the channel coefficients. In \eqref{eq:avg_var}, $(a)$ holds since $\hat{\mbf{x}}$ is an unbiased estimator, and $(b)$ follows from the fact that $J^{-1}$ is positive definite and $f(\mathcal{A})$ non-negative, which is sufficient for the integration to preserve the inequality. Note that $\bar{J}^{-1}(\mbf{x})$ is a $2\times 2$ matrix whose diagonal elements is a lower bound for the variance of each coordinate while the off-diagonal elements describes the error-covariance. To provide a scalar measure for the localization error, we can take the trace and square root on both sides of \eqref{eq:avg_var}, which gives
\begin{align}
    \sqrt{E[\norm{\hat{\mbf{x}} - \mbf{x}}^2]} \geq \sqrt{\mathbb{T}[\bar{J}^{-1}(\mbf{x})]} \triangleq \text{PEB}.
\end{align}

\ifCLASSOPTIONcaptionsoff
  \newpage
\fi

\bibliographystyle{IEEEtran}
\bibliography{references}

\end{document}